\documentclass[12pt]{article}
\usepackage{amssymb,amscd,array}
\catcode `\@=11
\@addtoreset{equation}{section}

\def\harr#1#2{\smash{\mathop{\hbox to .5in{\rightarrowfill}}
\limits^{\scriptstyle#1}_{\scriptstyle#2}}}

\def\harrl#1#2{\smash{\mathop{\hbox to .5in{\leftarrowfill}}
\limits^{\scriptstyle#1}_{\scriptstyle#2}}}

\def\qed{\blacksquare}
\newcommand{\be}{\begin{equation}}
\newcommand{\ee}{\end{equation}}
\newcommand{\bea}{\begin{eqnarray}}
\newcommand{\eea}{\end{eqnarray}}
\newcommand{\R}{\mathbb{R}}
\newcommand{\N}{\mathbb{N}}
\newcommand{\C}{\mathbb{C}}
\newcommand{\Z}{\mathbb{Z}}
\newtheorem{thm}{Theorem}[section]
\newtheorem{rem}[thm]{Remark}
\newtheorem{lemma}[thm]{Lemma}
\newtheorem{cor}[thm]{Corollary}
\newtheorem{prop}[thm]{Proposition}
\begin{document}
\begin{titlepage}
\begin{center}
{\bf \Large{Quantum Strings and Superstrings}}
\end{center}
\vskip 1.0truecm
\centerline{D. R. Grigore
\footnote{e-mail: grigore@theory.nipne.ro, grigore@ifin.nipne.ro}}
\vskip5mm
\centerline{Dept. of Theor. Phys., Inst. Atomic Phys.}
\centerline{Bucharest-M\u agurele, P. O. Box MG 6, ROM\^ANIA}
\vskip 2cm
\bigskip \nopagebreak
\begin{abstract}
\noindent
In the first sections of this paper we give an elementary but rigorous approach to the construction of the quantum Bosonic and supersymmetric string system continuing the analysis
of Dimock. This includes the construction of the DDF operators without using the vertex algebras. Next we give a rigorous proof of the equivalence between the light-cone and the covariant quantization methods. Finally, we provide a new and simple proof of the BRST quantization for these string models. 
\end{abstract}
\end{titlepage}

\section{Introduction}
We follow the analysis of Dimock \cite{Di1}, \cite{Di2} concerning the construction of the quantum Hilbert space of the Bosonic string and of the superstring the purpose being of presenting the facts as elementary as possible and in the same time rigorous. 

To be able to do that we start with a very simple method of constructing
representations of the Virasoro and Kac-Moody algebras acting in Bosonic
and Fermionic Fock spaces. We present a different way of computing things 
based on Wick theorem. Then we remind the main ingredients of the
light-cone formalism and we prove the Poincar\'e invariance of the
string and superstring systems. Most of the results obtained in this paper 
are known in the standard literature \cite{GRT}, \cite{GGRT}, \cite{GS},
\cite{GSW}, \cite{GH} but we offer some new simple proofs. There are various attempts to clarify the main mathematical aspects of this topics (see the references) but we are closest to the spirit of \cite{Lu}, \cite{Ot}, \cite{Ne} and \cite{Di1},\cite{Di2}. 

To establish the equivalence between the light-cone and the covariant formalism one needs the so-called DDF operators. Usually DDF operators are introduced using formal series which are elements of the so-called vertex algebras \cite{FB}. We present here an elementary derivation in the Bosonic case without using vertex algebras. Next we show that the covariant construction is equivalent (in the sense of group representation theory) with the light-cone construction; we are using the Hilbert space fiber-bundle formalism \cite{Va}. Finally we give an elementary proof of the BRST quantization procedure for all the string models considered previously. In particular we are able to find very explicit formulas for the cohomology of the BRST operator. 

\section{Quadratic Hamiltonians in Fock spaces \label{quad}}

\subsection{Bose systems of oscillators\label{bose}}

We consider the Hilbert space
${\cal H}$
with the scalar product
$<\cdot,\cdot>$
generated by $N$ Bose oscillators; the creation
and annihilation operators are  
$
a_{m},  a_{m}^{+}, \quad m = 1,\dots,N
$ 
and verify
\be
[ a_{m}, a_{n}^{+} ] = \delta_{mn} \cdot I \quad
[ a_{m}, a_{n} ] = 0, \quad [ a^{+}_{m}, a^{+}_{n} ] = 0,
\quad \forall m, n = 1,\dots,N.
\ee

If 
$
\Omega \in {\cal H}
$
is the vacuum state we have 
$
a_{m} \Omega = 0, \quad m > 0.
$
As usual \cite{GSW} it is more convenient to introduce the operators
$
\alpha_{m}, m = \{\pm 1,\dots,\pm N\}
$
according to:
\be
\alpha_{m} = \sqrt{m}~a_{m},  \forall m > 0 \qquad
\alpha_{m} = a_{-m}^{+}, \forall m < 0 
\ee

Then the canonical commutation relation from above can be compactly
written as follows:
\be
[ \alpha_{m}, \alpha_{n} ] = m~\delta_{m+n} \cdot I,
\quad \forall m, n \not= 0
\label{com-alpha}
\ee
where
$
\delta_{m} = \delta_{m,0}
$
and we also have
\bea
\alpha_{m} \Omega = 0, \quad m > 0
\nonumber \\
\alpha_{m}^{+} = \alpha_{-m} \quad \forall m.
\eea

To apply Wick theorem we will also need the $2$-point function; 
we easily derive
\be
<\Omega, \alpha_{m} \alpha_{n}\Omega> = \theta(m)~m~\delta_{m+n}, 
\quad \forall m, n \not= 0
\label{vac-alpha}
\ee
where
$\theta$
is the usual Heaviside function. The main result is the following
\begin{thm}
Let us consider operators of the form
\be
H(A) \equiv {1\over 2} A_{mn} :\alpha_{m}\alpha_{n}:
\ee
where $A$ is a symmetric matrix
$A^{T} = A$
and the double dots give the Wick ordering. Then:
\be
[ H(A), H(B) ] = H([A,B]) + \omega_{\alpha}(A,B) \cdot I
\label{hh-alpha}
\ee
where the commutator
$
[A,B] = A\cdot B - B\cdot A
$
is computed using the following matrix product
\be
(A\cdot B)_{pq} \equiv \sum_{m \not= 0} m~A_{pm}~B_{-m,q}
\label{multi}
\ee 
and we have defined
\be
\omega_{\alpha}(A,B) \equiv {1\over 2} \sum_{m,n > 0} mn~A_{mn}~B_{-n,-m} 
- (A \leftrightarrow B).
\ee
\label{H-alpha}
\end{thm}

{\bf Proof:} Is elementary and relies on computing the expression
$H(A)H(B)$
using Wick theorem:
\bea
H(A) H(B) = :H(A)H(B): 
\nonumber \\
+ {1\over 4} A_{mn} B_{pq} [ <\Omega, \alpha_{m}\alpha_{p}\Omega> 
:\alpha_{n}\alpha_{q}: + (m \leftrightarrow n) +
(p \leftrightarrow q) + (m \leftrightarrow n, p \leftrightarrow q) ] 
\nonumber \\
+ <\Omega, \alpha_{m}\alpha_{p}\Omega> <\Omega, \alpha_{n}\alpha_{q}\Omega> \cdot I
+ <\Omega, \alpha_{m}\alpha_{q}\Omega> <\Omega, \alpha_{n}\alpha_{p}\Omega> \cdot I]
\eea

If we use the $2$-point function we easily arrive at the formula from the 
statement.
$\qed$

Now we extend the previous result to the case when we have an infinite number
of oscillators. We consider that 
$m \in \Z^{*}$
and the matrix $A$ is {\it semi-finite} i.e. there exists 
$N > 0$ 
such that
\be
A_{mn} = 0 \quad {\rm for} \quad |m + n| > N.
\label{semi}
\ee

We note that if $A$ and $B$ are semi-finite then
$A \cdot B$
is also semi-finite. We will need the {\it algebraic Fock space} which is the subspace
${\cal D}_{0} \subset {\cal H}$
with finite number of particles. The elements of 
${\cal D}_{0}$
are, by definition, finite linear combinations of vectors of the type
$
a^{+}_{m_{1}}\dots a^{+}_{m_{k}}\Omega;
$
the subspace
${\cal D}_{0}$
is dense in
${\cal H}$.
Then one can prove easily that the operator
$H(A)$
is well defined on
${\cal D}_{0}$,
leaves
${\cal D}_{0}$
invariant and formula (\ref{hh-alpha}) remains true in 
${\cal D}_{0}$.

We will need an extension of this result namely we want to consider the case when
the index $m$ takes the value 
$m = 0$
also i.e.
$m \in \Z$
and we preserve the commutation relation (\ref{com-alpha}). 
We note that the relation (\ref{vac-alpha}) is not valid if one (or both) of the indices are null so the previous proof does not
work. It can be proved, however, directly that the statement of the theorem remains
true if we extend accordingly the definition of the matrix product including the 
value $0$ also i.e in (\ref{multi}) we leave aside the restriction
$m = 0$.
In general, the Hilbert space of this case will not be entirely
of Fock type: the operators
$\alpha_{m}~m \not= 0$
will live in a Fock space tensored with another Hilbert space where live
the operators
$\alpha_{0}$.
\subsection{A Systems of Fermi Oscillators\label{fermi-b}}

We consider the Hilbert space
${\cal H}$
with the scalar product
$<\cdot,\cdot>$
generated by $N$ Fermi oscillators; the creation
and annihilation operators are  
$
b_{m},  b_{m}^{+}, \quad m = 1,\dots,N
$ 
and verify
\be
\{ b_{m}, b_{n}^{+} \} = \delta_{mn} \cdot I \quad
\{ b_{m}, b_{n} \} = 0, \quad \{ b^{+}_{m}, b^{+}_{n} \} = 0, 
\quad \forall m, n \not= 0.
\ee

If 
$
\Omega \in {\cal H}
$
is the vacuum state we have 
$
b_{m} \Omega = 0, \quad m > 0.
$
As above it is more convenient to introduce the operators
$
b_{m}, m = \{\pm 1,\dots,\pm N\}
$
according to:
\be
b_{m} =  b_{m}, \forall m > 0 \qquad
b_{m} = b_{-m}^{+}, \forall m < 0
\ee
and the canonical anti-commutation relation from above can be compactly
rewritten as follows:
\be
\{ b_{m}, b_{n} \} = \delta_{m+n} \cdot I,
\quad \forall m, n \not= 0.
\label{com-b}
\ee

We also have
\bea
b_{m} \Omega = 0, \quad m > 0
\nonumber \\
b_{m}^{+} = b_{-m}.
\eea

The $2$-point function is in this case:
\be
<\Omega, b_{m} b_{n}\Omega> = \theta(m) \delta_{m+n}, 
\quad \forall m, n \not= 0.
\label{vac-b}
\ee

The main result is the following
\begin{thm}
Let us consider operators of the form
\be
H(A) \equiv {1\over 2} A_{mn} :b_{m}b_{n}:
\ee
where $A$ is a antisymmetric matrix
$A^{T} = - A$
and the double dots give the Wick ordering. Then:
\be
[ H(A), H(B) ] = H([A,B]) + \omega_{b}(A,B) \cdot I
\ee
where the commutator
$
[A,B] = A\cdot B - B\cdot A
$
is computed using the following matrix product
\be
(A\cdot B)_{pq} \equiv \sum_{m \not= 0} A_{pm} B_{-m,q}
\ee 
and we have defined
\be
\omega_{b}(A,B) \equiv {1\over 2} \sum_{m,n > 0} A_{mn} B_{-n,-m} 
- (A \leftrightarrow B).
\ee
\label{H-b}
\end{thm}

The proof is similar to the preceding theorem. The previous result can be extended 
to the case when we have an infinite number of oscillators i.e. 
$m \in \Z^{*}$
and the matrix $A$ is semi-finite: the operator
$H(A)$
is well defined on the corresponding algebraic Fock space
${\cal D}_{0}$,
leaves invariant this subspace and the previous theorem remains true.

\subsection{Another System of Fermi Oscillators\label{fermi-d}}

We extend the previous results to the case when the value
$m = 0$
is allowed i.e. the Hilbert space 
$
{\cal H}
$
is generated by the operators  
$
d_{m},  \quad m = -N,\dots,N
$ 
and verify
\bea
\{ d_{m}, d_{n} \} = \delta_{mn} \cdot I \quad \forall m, n
\nonumber \\
d_{m}^{\dagger} = d_{-m}\quad \forall m
\nonumber \\
d_{m}\Omega = 0 \quad \forall m > 0
\label{dd}
\eea
where
$
\Omega \in {\cal H}
$
is the vacuum state. One can realize this construction if one takes
$
{\cal H} = {\cal F} \otimes^{s} {\cal C}
$
where
$
{\cal F}
$
is the Fock space from the preceding Section,
$
{\cal C}
$
is the Clifford algebra generated by the element
$
b_{0}
$
verifying
$
b_{0}^{\dagger} = b_{0} \qquad b_{0}^{2} = 1/2
$
and the skew tensor product
$
\otimes^{s}
$
is chosen such that the operators
\be
d_{n} \equiv b_{n} \otimes^{s} I_{2} \quad  \forall m \not= 0 \qquad 
d_{0} \equiv I_{1} \otimes^{s} b_{0}
\ee
verify (\ref{dd}). Another more explicit construction is to consider the Hilbert space is generated by the creation and annihilation operators  
$
b_{m},  b_{m}^{+}, \quad m = 0,\dots,N
$  
such that we have 
$
b_{m} \Omega = 0, \quad m \geq 0
$
and to define the operators
$
d_{m}, m = \{-N,\dots,N\}
$
according to:
\bea
d_{m} = \cases{ b_{m}, & for m $>$ 0 \cr
{1\over \sqrt{2}} (b_{0} + b_{0}^{+}), & for m = 0 \cr
b_{-m}^{+}, & for m $<$ 0 \cr}
\label{d-b}
\eea

The $2$-point function is in this case:
\be
<\Omega, d_{m} d_{n}\Omega> = \theta_{+}(m) \delta_{m+n}, 
\quad \forall m, n
\label{vac-d}
\ee
where we have introduced the modified Heaviside function
\bea
\theta_{+}(m) = \cases{ 1, & for m $>$ 0 \cr
{1\over 2}, & for m = 0 \cr
0, & for m $<$ 0 \cr}.
\eea

It follows that the main result is similar to the previous one: 
\begin{thm}
Let us consider operators of the form
\be
H(A) \equiv {1\over 2} A_{mn} :d_{m}d_{n}:
\ee
where $A$ is a antisymmetric matrix
$A^{T} = - A$
and the double dots give the Wick ordering. Then:
\be
[ H(A), H(B) ] = H([A,B]) + \omega_{d}(A,B) \cdot I
\ee
where the commutator
$
[A,B] = A\cdot B - B\cdot A
$
is computed using the following matrix product
\be
(A\cdot B)_{pq} \equiv \sum_{m = -N}^{N} A_{pm} B_{-m,q}
\ee 
and we have defined
\be
\omega_{d}(A,B) \equiv {1\over 2} \sum_{m,n \geq 0} \theta_{+}(m) \theta_{+}(n) 
A_{mn} B_{-n,-m} 
- (A \leftrightarrow B).
\ee
\label{H-d}
\end{thm}

The proof is similar to the preceding Section. The previous result can be extended 
to the case when we have an infinite number of oscillators i.e. 
$m \in \Z$
and the matrix $A$ is semi-finite: the operator
$H(A)$
is well defined on the corresponding algebraic Fock space
${\cal D}_{0}$,
leaves invariant this subspace and the previous theorem remains true.

\begin{rem}
It follows easily that the expressions
$
\omega_{\alpha},\omega_{b},\omega_{d}
$
are $2$-cocyles. They are quantum obstructions (or anomalies) because they do not appear
if we work in classical field theory replacing the commutators by Poisson brackets.
\end{rem}

\section{Virasoro Algebras in Fock Spaces\label{vir}}

We have constructed some Fock spaces for which we have a nice commutation relation
of the bilinear operators. In all these cases we will be able to construct 
representations of the Virasoro algebra taking convenient expressions for the
matrix $A$. We give the details corresponding to the structure of the closed
strings.
\subsection{Bose Case\label{vir-bose}}

\begin{thm}
In the conditions of theorem \ref{H-alpha} the operators given by the formulas
\be
L_{m} \equiv {1\over 2}~\sum_{n \not= 0,m}~:\alpha_{m-n} \alpha_{n}: 
\label{vir-alpha}
\ee
are well defined on the algebraic Fock subspace, leave invariant this subspace and verify the following relations:
\bea
[ L_{m}, L_{n} ] = (m - n) L_{m+n} + {m (m^{2} - 1) \over 12}~\delta_{m+n}~\cdot I
\nonumber \\
L_{m}^{+} = L_{-m} \quad \forall m \in \Z
\nonumber \\
L_{0} \Omega = 0.
\eea
\end{thm}
{\bf Proof:} We consider the matrices
$A_{m}$
given by
$
(A_{m})_{pq} \equiv \delta_{p+q-m}
$
and we are in the conditions of theorem \ref{H-alpha}: the matrices
$A_{m}$
are symmetric and semi-finite. It remains to prove that
$
[ A_{m}, A_{n} ] = (m -n) A_{m+n}
$
and to compute the $2$-cocycle
$\omega_{\alpha}(A_{m}, A_{n}) = {m(m^{2} - 1) \over 12} \delta_{m+n}$
and we obtain the commutation relation from the statement.
$\qed$
 
One can express everything in terms of the original creation and annihilation operators
$a_{m}^{\#}$;
if we use the holomorphic representation for the harmonic oscillator operators 
$
a_{m}^{+} = z_{m} \quad a_{m} = {\partial \over \partial z_{m}}
$
we obtain the formula (7.2.10) from \cite{Ne}.

It is important that we can extend the previous results to the case when
$\alpha_{0} \not= 0$
(see the end of Subsection \ref{bose}.) To preserve (\ref{com-alpha}) we impose
\be
[ \alpha_{0}, \alpha_{m} ] = 0 \quad \forall m \in \Z^{*} 
\ee 
and we keep the relation (\ref{vir-alpha}) without the restrictions 
$n \not= 0, m$; explicitly:
\be
L_{m} = \dots + \alpha_{m} \alpha_{0} \quad \forall m \not= 0,
\qquad
L_{0} = \dots + {1\over 2}~\alpha_{0}^{2}
\ee
where by $\dots$ we mean the expressions from the preceding theorem corresponding to 
$\alpha_{0} \equiv 0$.
Eventually we have to consider a larger Hilbert space containing as a subspace the Fock space generated by the operators
$\alpha_{n}\quad  n \not= 0$.
By direct computations we can prove that the statement of the theorem remains true; also
we have
\be
[ L_{m}, \alpha_{n} ] = - n~\alpha_{m+n}.
\label{l-a}
\ee
In the following we will use only the case when
$\alpha_{0} \not= 0$. 

\subsection{First Fermi Case\label{vir-ns}}

We have a similar result for the Fermi operators of type $b$: we will consider
that these operators are 
$b_{r}$
indexed by 
$r \in {1\over 2} + \Z$
and they verify:
\bea
\{ b_{r}, b_{s} \} = \delta_{r+s} \cdot I,
\quad \forall r,s \in {1\over 2} + \Z
\nonumber \\
b_{r} \Omega = 0, \quad r > 0
\nonumber \\
b_{r}^{+} = b_{-r}.
\eea
Then:
\begin{thm}
In the conditions of theorem \ref{H-b} the operators given by the formulas
\be
L_{m} \equiv {1\over 2}~\sum_{r \in 1/2 + \Z}~\left(r + {m\over 2}\right)
:b_{-r} b_{m+r}: = {1\over 2}~\sum_{r \in 1/2 + \Z}~r~:b_{-r} b_{m+r}:
\label{vir-b}
\ee
are well defined on the algebraic Fock subspace, leave invariant this subspace and verify the following relations:
\bea
[ L_{m}, L_{n} ] = (m - n) L_{m+n} + {m (m^{2} - 1) \over 24}~\delta_{m+n}~\cdot I. 
\nonumber \\
~[ L_{m}, b_{r} ] = - \left(r + {m\over 2}\right)~b_{m+r}
\nonumber \\
L_{m}^{+} = L_{-m}
\nonumber \\
L_{0} \Omega = 0.
\eea
\end{thm}
{\bf Proof:} We consider the matrices
$A_{m}$
given by
$
(A_{m})_{rs} \equiv {1\over 2} (s - r) \delta_{r+s-m}
$
and we are in the conditions of theorem \ref{H-b}: the matrices
$A_{m}$
are anti-symmetric and semi-finite. It remains to compute the $2$-cocycle
$\omega_{b}$
to obtain the commutation relation from the statement.
$\qed$
 
If we use the representation in terms of Grassmann variables 
$
b_{r}^{+} = \xi_{r} \quad b_{r} = {\partial \over \partial \xi_{r}}
$
for these operators we obtain the formulas from \cite{Ne} pg. 225.
 
\subsection{Second Fermi Case\label{vir-r}}

Finally we have a similar result for the Fermi operators of type $d$.
\begin{thm}
In the conditions of theorem \ref{H-d} the operators given by the formulas
\be
L_{m} \equiv {1\over 2}~\sum_{n \in \Z}~\left(n + {m\over 2}\right)
:d_{-n} d_{m+n}: = {1\over 2}~\sum_{n \in \Z}~n~:d_{-n} d_{m+n}:
\label{vir-d}
\ee
are well defined on the algebraic Fock subspace, leave invariant this subspace and verify the following relations:
\bea
[ L_{m}, L_{n} ] = (m - n) L_{m+n} + {m (m^{2} + 2) \over 24}~\delta_{m+n}~\cdot I. \nonumber \\
~[ L_{m}, d_{n} ] = - \left(n + {m\over 2}\right)~b_{m+n}
\nonumber \\
L_{m}^{+} = L_{-m}
\nonumber \\
L_{0} \Omega = 0.
\eea
\end{thm}
{\bf Proof:} We consider the matrices
$A_{m}$
given by
$
(A_{m})_{pq} \equiv {1\over 2} (q - p) \delta_{p+q-m}
$
and we are in the conditions of theorem \ref{H-d}: the matrices
$A_{m}$
are anti-symmetric and semi-finite. It remains to compute the $2$-cocycle
$\omega_{d}$
to obtain the commutation relation from the statement.
$\qed$
 
We observe that in the commutation relation of the preceding theorem the expression
of the cocycle is different from the usual form
$
c~{m(m^{2} - 1)\over 12};
$
we can fix this inconvenience immediately if we define:
\be
\tilde{L}_{m} \equiv L_{m} \quad \forall m \not= 0 
\qquad
\tilde{L}_{0} \equiv L_{0} + {1\over 16} \cdot I;
\ee 
we obtain in this case:
\be
[ \tilde{L}_{m}, \tilde{L}_{n} ] = (m - n) \tilde{L}_{m+n} 
+ {m (m^{2} - 1) \over 24}~\delta_{m+n}~\cdot I. 
\ee
and
\be
\tilde{L}_{0} \Omega = {1\over 16} \cdot I.
\ee

\subsection{Multi-Dimensional Cases\label{vir-multi}}

In the preceding Subsections we have obtained three representations of the Virasoro algebra corresponding to
$
(c,h) = (1, 0), \left({1\over 2}, 0\right), \left({1\over 2}, {1\over 16} \right). 
$

The previous results can be easily extended to a more general case. Let
$
\eta^{jk} = \eta_{jk}, \quad j,k = 1,\dots,D
$
be a diagonal matrix with the diagonal elements 
$
\epsilon_{1},\dots,\epsilon_{D} = \pm 1. 
$

In the Bose case we can consider that we have the family of operators:
$
\alpha^{j}_{m}, m \in \Z, j = 1,\dots,D
$
acting in the Hilbert space
$
{\cal F}^{(\alpha)}
$
such that:
\bea
[ \alpha^{j}_{m}, \alpha^{k}_{n} ] = m~\eta_{jk}\delta_{m+n} \cdot I, 
\quad \forall m, n
\nonumber \\
\alpha^{j}_{m} \Omega = 0, \quad m > 0
\nonumber \\
(\alpha^{j}_{m})^{+} = \alpha^{j}_{-m} \quad \forall m.
\eea

We can define
\be
L^{(\alpha)}_{m} \equiv {1\over 2}~\sum_{n \in \Z}~\eta_{jk}:\alpha^{j}_{m-n} \alpha^{k}_{n}: 
\label{vir-alpha-D}
\ee
and we have:
\be
[ L^{(\alpha)}_{m}, L^{(\alpha)}_{n} ] = (m - n) L^{(\alpha)}_{m+n} 
+ D~{ m (m^{2} - 1) \over 12}~\delta_{m+n}~\cdot I. 
\ee

In the first Fermi case we have the operators:
$
b^{j}_{r}, r \in {1\over 2} + \Z, j = 1,\dots,D
$
acting in the Hilbert space
$
{\cal F}^{(b)}
$
such that:
\bea
\{ b^{j}_{r}, b^{k}_{s} \} = \eta_{jk}~\delta_{r+s} \cdot I,
\quad \forall r,s
\nonumber \\
b^{j}_{r} \Omega = 0, \quad r > 0
\nonumber \\
(b^{j}_{r})^{+} = b^{j}_{-r} \quad \forall r.
\eea
We define
\be
L^{(b)}_{m} \equiv {1\over 2}~\sum_{r \in 1/2 + \Z}~
\left(r + {m\over 2}\right)~\eta_{jk}
:b^{j}_{-r} b^{k}_{m+r}: 
\label{vir-b-D}
\ee
are well defined and verify the following relations:
\be
[ L^{(b)}_{m}, L^{(b)}_{n} ] = (m - n) L^{(b)}_{m+n} 
+ D~{m (m^{2} - 1) \over 24}~\delta_{m+n}~\cdot I. 
\ee

Finally, in the second Fermi case we have the operators
$
d^{j}_{m}, m \in \Z, j = 1,\dots,D
$
acting in the Hilbert space
$
{\cal F}^{d}
$
such that
\bea
\{ d^{j}_{m}, d^{k}_{n} \} = \eta_{jk}~\delta_{m+n} \cdot I
\nonumber \\
d^{j}_{m} \Omega = 0, \quad m > 0
\nonumber \\
(d^{j}_{m})^{+} = d^{j}_{-m}\quad \forall m.
\eea
We can define
\be
L^{(d)}_{m} \equiv {1\over 2}~\sum_{n \in \Z}~\left(n + {m\over 2}\right)~\eta_{jk}
:d^{j}_{-n} d^{k}_{m+n}: 
\label{vir-d-D}
\ee
and we have the following relations:
\be
[ L^{(d)}_{m}, L^{(d)}_{n} ] = (m - n) L^{(d)}_{m+n} 
+ D~{m (m^{2} + 2) \over 24}~\delta_{m+n}~\cdot I. 
\ee

We redefine
\be
\tilde{L}^{(d)}_{m} \equiv L^{(d)}_{m} \quad \forall m \not= 0 
\qquad
\tilde{L}^{(d)}_{0} \equiv L^{(d)}_{0} + {D\over 16} \cdot I;
\label{shift}
\ee 
we obtain in this case:
\be
[ \tilde{L}^{(d)}_{m}, \tilde{L}^{(d)}_{n} ] = (m - n) \tilde{L}^{(d)}_{m+n} 
+ D~{m (m^{2} - 1) \over 24}~\delta_{m+n}~\cdot I. 
\ee
\be
\tilde{L}^{(d)}_{0} \Omega = {D\over 16} \cdot I.
\ee

In all these cases the $2$-cocycle gets multiplied by $D$. The Hilbert space has a
positively defined scalar product only in the case
$
\epsilon_{1} = \cdots = \epsilon_{D} = 1.
$

We can combine the Bose and Fermi cases as follows. We consider the Hilbert spaces
$
{\cal F}^{(NS)} \equiv {\cal F}^{(\alpha)} \otimes {\cal F}^{(b)}
$
and
$
{\cal F}^{(R)} \equiv {\cal F}^{(\alpha)} \otimes {\cal F}^{(d)}
$
respectively; the Virasoro operators are
\bea
L^{(NS)}_{m} \equiv L^{(\alpha)} \otimes I_{2} + I_{1} \otimes L^{(b)}
\nonumber \\
L^{(R)}_{m} \equiv L^{(\alpha)} \otimes I_{2} + I_{1} \otimes \tilde{L}^{(d)} 
\eea
and we have in both cases:
\be
[ L^{(NS,R)}_{m}, L^{(NS,R)}_{n} ] = (m - n) L^{(NS,R)}_{m+n} 
+ D~{m (m^{2} - 1) \over 8}~\delta_{m+n}~\cdot I. 
\label{lll}
\ee

These two constructions are called {\it Neveau-Schwartz} (and {\it Ramond}) respectively.
In these cases one can extend the Virasoro algebra to a {\it super-Virasoro} algebra 
\cite{GSW}.

We conclude this Subsection with some simple propositions. First we have a natural
representation of the rotation group in the Fock space:
\begin{prop}
Suppose that the signature of $\eta$ is
$(r,s)$;
then we can define in the corresponding Hilbert spaces a representation of the Lie
algebra
$so(r,s)$
according to:
\bea
J^{(\alpha)jk} \equiv - i~\sum_{m > 0} {1\over m}
\alpha_{-m}^{j} \alpha_{m}^{k} - (j \leftrightarrow k)
\nonumber \\
J^{(b)jk} \equiv - i~\sum_{r > 0} b_{-r}^{j} b_{r}^{k} - (j \leftrightarrow k)
\nonumber \\
J^{(d)jk} \equiv - i~\sum_{m > 0} d_{-m}^{j} d_{m}^{k} - (j \leftrightarrow k)
\label{j}
\eea
respectively. 
\end{prop}
Indeed, we can obtain directly from the (anti)commutation relations in all the cases:
\be
[ J^{kl}, J^{pq} ] = - i~( \eta^{lp}~J^{jq} + \eta^{jq}~J^{lp} 
- \eta^{kp}~J^{lq} + \eta^{lq}~J^{kp}).
\ee

We also note that the Virasoro operators are rotational invariant: in all cases
\be
[ J^{kl} , L_{m} ] = 0.
\ee

Next, we have a proposition which will be important for proving the Poincar\'e
invariance:
\begin{prop}
If
$\Psi \in {\cal D}_{0}$
is an arbitrary vector from the algebraic Hilbert space then we have in all cases
\be
L_{m} \Psi = 0 
\ee 
for sufficiently large 
$m > 0$.
\label{aaP}
\end{prop}
{\bf Proof:} We consider only the one-dimensional Bose case; the other cases are similar. If
$\Psi$ 
is a vector in the algebraic Fock space it is clear that we have
\be
\alpha_{m} \Psi = 0
\label{aP}
\ee
for $m$ sufficiently large. This implies immediately that all the sums in
(\ref{vir-alpha}) are finite (it is better to re-express everything in terms of the original creation and annihilation operators). It is clear that 
$
\sum_{n > 0} \dots a^{+}_{n} a_{m+n} \Psi = 0
$
for sufficiently large $m$ because of (\ref{aP}). Also
$
\sum_{n = 1}^{m-1} \dots a_{n} a_{m-n} \Psi = 0
$
for sufficiently large $m$; indeed, all the indices in the preceding sum are
larger than 
$
{m\over 2} 
$
so again we can apply (\ref{aP}).
$\qed$
 
It is known that any $2$-cocycle of the Virasoro algebra is cohomologous to a standard form
$
c~{m(m^{2} - 1)\over 12};
$
however, we can always add a trivial cocycle.
\begin{prop}
Suppose that we have
\be
[ L_{m}, L_{n} ] = (m - n) L_{m+n} + c~{m (m^{2} - 1) \over 12}~\delta_{m+n}~\cdot I 
\ee
and we redefine 
\be
L_{m}(a) \equiv L_{m} \quad \forall m \not= 0 
\qquad
L_{0}(a) \equiv L_{0} - a \cdot I; \qquad a \in \R.
\ee 
Then we have:
\be
[ L_{m}(a), L_{n}(a) ] = (m - n) L_{m+n}(a) 
+ \left[ c~{m (m^{2} - 1) \over 12} + 2 a m \right]~\delta_{m+n}~\cdot I. 
\label{ll-c}
\ee
\end{prop}

\section{Kac-Moody Algebras in Fock Spaces\label{km}}

We use here the techniques from Section \ref{quad}. We suppose that we have
$
t_{A}, A = 1,\dots,r
$
a $N$-dimensional representation of the Lie algebra
$\mathfrak{g}$:
\be
[ t_{A}, t_{B} ] = f_{ABC} t_{C};
\ee
here
$
f_{ABC}
$
are the {\it structure constants} of 
$\mathfrak{g}$
and we need also the {\it Killing-Cartan form} of the representation $t$:
\be
g_{AB} \equiv Tr(t_{A}t_{B}).
\ee

We will suppose that the matrices
$
t_{A}
$
are antisymmetric:
\be
t_{A}^{T} = - t_{A}.
\ee

Finally, we will need the {\it contra-gradient} representation
\be
\tilde{t}_{A} = - t^{+}_{A} = \bar{t}_{A} \quad \forall A.
\ee

We will construct some representations of the Kac-Moody algebras acting in
${\cal F}^{(b)}$
and
${\cal F}^{(d)}$
respectively.
\subsection{Neveau-Schwartz Case}
In this case
\begin{thm}
In the conditions of theorem \ref{H-b} the operators given by the formulas
\be
K^{A}_{m} \equiv {1\over 2}~(t_{A})_{jk}~\sum_{r \in 1/2 + \Z}~:b^{j}_{-r} b^{k}_{m+r}: 
\label{km-b}
\ee
are well defined on the algebraic Fock subspace, leave invariant this subspace and verify the following relation:
\be
[ K^{A}_{m}, K^{B}_{n} ] = f_{ABC} K^{C}_{m+n} + {1 \over 2}~m~g_{AB}\delta_{m+n}~\cdot I. 
\label{kk-ns}
\ee

We have in this case also the Hermiticity property:
\be
(K^{A}_{m})^{+} = - \tilde{K}^{A}_{-m}
\label{herm-km-b}
\ee
where 
$
\tilde{K}^{A}_{m}
$
is associated to the representation
$
\tilde{t}_{A}.
$

The following commutation relation are true:
\be
[ K^{A}_{m}, L_{n} ] = m~K^{A}_{m+n}.
\label{kl-b}
\ee

If
$
\Psi \in {\cal D}_{0}
$
then we have
\be
K^{A}_{m} \Psi = 0
\label{kp-ns}
\ee
for 
$m > 0$
large enough.
\end{thm}
{\bf Proof:} We consider the matrices
$A^{A}_{m}$
given by
$
A^{A}_{m} \equiv \tilde{A}_{m} \otimes t_{A}
$
where
$
(\tilde{A}_{m})_{rs} \equiv \delta_{r+s-m}
$
and we are in the conditions of theorem \ref{H-b}: the matrices
$A^{A}_{m}$
are anti-symmetric (because
$\tilde{A}_{m}$
are symmetric and
$t_{A}$
are anti-symmetric) and semi-finite. We can easily prove that
$
[ A^{A}_{m}, A^{B}_{n} ] = f_{ABC} A^{C}_{m+n};
$
it remains to compute the $2$-cocycle
$\omega_{b}$
to obtain the commutation relation (\ref{kk-ns}). The relation (\ref{kl-b})
follows from the same theorem \ref{H-b} if we take 
$
A \rightarrow A_{m}^{A}, \quad  B \rightarrow A_{n} \otimes I_{2};
$
in this case the cocycle
$\omega_{b}$
is null because 
$
Tr(t_{A}) = 0.
$
The relation (\ref{kp-ns}) is proved as Proposition \ref{aaP}.
$\qed$

We have an important particular case:
\begin{prop}
Let us consider the case 
$\mathfrak{g} = so(D)$
with the fundamental representation
\be
(E_{jk})_{pq} = \delta_{jp} \delta_{kq} - \delta_{jq} \delta_{kp}. 
\ee

Then the associated operators, denoted by
$
K^{jk}_{m}
$
verify the following relations:
\be
[ K^{jk}_{m}, K^{pq}_{n} ] = \delta_{kp} K^{jq}_{m+n} - \delta_{kq} K^{jp}_{m+n}
+ \delta_{jq} K^{kp}_{m+n} - \delta_{jp} K^{kq}_{m+n}
+ m~( \delta_{kp} \delta_{jq} - \delta_{kq} \delta_{jp})~\delta_{m+n} \cdot I
\label{kkk}
\ee
\be
(K^{jk}_{m})^{+} = - K^{jk}_{-m}
\label{herm-km-b-D}
\ee
\end{prop}
{\bf Proof:}
We apply theorem \ref{H-b}; in our case we have by direct computation
\be
[ E_{jk}, E_{pq} ] = \delta_{kp} E_{jq} - \delta_{kq} E_{jp} + 
\delta_{jq} E_{kp} - \delta_{jp} E_{kq}
\ee
and the Killing-Cartan form
\be
g_{jk,pq} = 2 ( \delta_{kp} \delta_{jq} - \delta_{kq} \delta_{jp}).
\ee
Then we apply the preceding theorem.
$\qed$

From the commutation relation of the preceding Proposition we observe that
- see (\ref{j}):
$
J^{(b)jk} = - i~K^{jk}_{0}
$
so we have another proof that these operators give a representation of the group
$
so(D).
$

Finally we give the following technical result \cite{GS}:
\begin{prop}
In the preceding conditions, the following formula is true
\bea
\sum_{m > 0} m~[ K^{jl}_{-m} K^{kl}_{m} - (j \leftrightarrow k) ] 
+ 2 \sum_{m > 0} ( K^{jk}_{-m} L_{m} + L_{-m}K^{jk}_{m}) 
+ 2 K^{jk}_{0} L_{0}
\nonumber \\
= {8 - D \over 8} \sum_{r > 0} (4 r^{2} - 1) [ :b^{j}_{-r} b^{k}_{r}:
- (j \leftrightarrow k) ] + ( 1 - 2a )~K^{jk}_{0};
\label{kkl-ns}
\eea
here
$
L_{m} = L_{m}(a).
$
\end{prop}
{\bf Proof:} First we note that all the sums are in fact finite when applied on the
algebraic Fock space. Then one substitutes the expressions for $K$'s and $L$'s
and applies Wick theorem. The computation is tedious but straightforward.
$\qed$

Let us note that we have a quantum anomaly in the right-hand side which vanishes
for
$
D = 8, \quad a = {1\over 2}.
$
\subsection{Ramond Case}

\begin{thm}
In the conditions of theorem \ref{H-d} the operators given by the formulas
\be
K^{A}_{m} \equiv {1\over 2}~(t_{A})_{jk}~\sum_{n \in \Z}~:d^{j}_{-n} d^{k}_{m+n}: 
\label{km-d}
\ee
are well defined on the algebraic Fock subspace, leave invariant this subspace and verify the same relations as in the preceding theorem.
\end{thm}
{\bf Proof:} We consider the matrices
$A_{m}$
given by
$
A^{A}_{m} \equiv \tilde{A}_{m} \otimes t_{A}
$
where
$
(A_{m})_{pq} \equiv \delta_{p+q-m}
$
and we are in the conditions of theorem \ref{H-d}.
$\qed$

The technical result is in this case:
\begin{prop}
In the preceding conditions, the following formula is true
\bea
\sum_{m > 0} m~[ K^{jl}_{-m} K^{kl}_{m} - (j \leftrightarrow k) ] 
+ 2 \sum_{m > 0} ( K^{jk}_{-m} L_{m} + L_{-m}K^{jk}_{m})
+ 2 K^{jk}_{0} L_{0}
\nonumber \\
= {8 - D \over 2} \sum_{k > 0} k^{2} [ :d^{j}_{-r} d^{k}_{r}:
- (j \leftrightarrow k) ] + \left( {D \over 8} - 2a\right)~K^{jk}_{0}.
\label{kkl-r}
\eea
\end{prop}

Again we have in the right-hand side a quantum anomaly which vanishes for
$
D = 8, \quad a = {1\over 2}.
$ 
\begin{rem}
We note that in the relations (\ref{kkl-ns}) 
and (\ref{kkl-r}) the appearance of the constant $a$ is in fact spurious: if we
express the Virasoro operators
$
L_{m}
$
in terms of the operators
$\alpha, b$
or $d$ then the constant $a$ drops out.
\end{rem}

\section{Light-Cone Coordinates}
In this Section we remind the basic ingredients of the light-cone description of
the representations of the Poincar\'e group. We consider this group in $D$ dimensions
and we denote the indices by
$
\mu, \nu = 0,1,\dots,D-1
$.
Let
$
\eta^{\mu\nu}
$
be the Minkowski diagonal form with
$
diag(\eta) = 1,-1,\dots,-1.
$

Then the algebra of the Poincar\'e group is generated by the basis elements
$
P^{\mu},\quad J^{\mu\nu}
$
where
$
J^{\mu\nu}
$
is anti-symmetric. The Lie brackets are:
\bea
~[ P^{\mu}, P^{\nu} ] = 0
\nonumber \\
~[ P^{\mu}, J^{\nu\rho} ] = i~\eta^{\mu\nu}~P^{\rho} - i~\eta^{\mu\rho}~P^{\nu}
\nonumber \\
~[ J^{\mu\nu}, J^{\rho\lambda} ] 
=  i~\eta^{\nu\rho}~J^{\mu\lambda} - i~\eta^{\mu\rho}~J^{\nu\lambda} 
- i~\eta^{\nu\lambda}~J^{\mu\rho} + i~\eta^{\mu\lambda}~J^{\nu\rho}.
\label{p1}
\eea

Let us change the basis
$
(P^{\mu}, J^{\mu\nu}) \rightarrow (P^{\pm}, P^{j}, J^{+-}, J^{j\pm}, J^{jk})
$
where
$j = 1,\dots,D-2$
and
\bea
P^{\pm} \equiv {1\over \sqrt{2}} (P^{0} \pm P^{D-1})
\nonumber \\
J^{j\pm} \equiv {1\over \sqrt{2}} (J^{j0} \pm J^{j,D-1})
\nonumber \\
J^{+-} \equiv J^{0,D-1}.
\eea

Then in the new basis the Lie brackets are:
\bea
~[ P^{\epsilon_{1}}, P^{\epsilon_{2}} ] = 0 \quad
[ P^{\epsilon}, P^{j} ] = 0 \quad
[ P^{j}, P^{k} ] = 0
\nonumber \\
~[ P^{\epsilon}, J^{jk} ] = 0 \quad
[ P^{j}, J^{+-} ] = 0
\nonumber \\
~[ P^{\epsilon}, J^{+-} ] = i~\epsilon~P^{\epsilon}
\nonumber \\
~[ P^{\epsilon_{1}}, J^{j\epsilon_{2}} ] = - {i \over 2} (1 - \epsilon_{1}\epsilon_{2})~P^{j} 
\nonumber \\
~[ P^{j}, J^{kl} ] = - i~\delta_{jk}~P^{l} + i~\delta_{jl}~P^{k}
\nonumber \\
~[ P^{j}, J^{k\epsilon} ] = - i~\delta_{jk}~P^{\epsilon}
\nonumber \\
~[ J^{+-}, J^{jk} ] = 0 
\nonumber \\
~[ J^{+-}, J^{j\epsilon} ] =  - i~\epsilon~J^{j\epsilon} 
\nonumber \\
~[ J^{j\epsilon_{1}}, J^{k\epsilon_{2}} ] 
=  {i\over 2}~(-1 +\epsilon_{1}\epsilon_{2})~J^{jk} 
+ {i\over 2}~(\epsilon_{2} - \epsilon_{1})~\delta_{jk}~J^{+-} 
\nonumber \\
~[ J^{j\epsilon}, J^{kl} ] 
=  - i~\delta_{jk}~J^{l\epsilon} + i~\delta_{jl}~J^{k\epsilon} 
\nonumber \\
~[ J^{kl}, J^{pq} ] 
=  - i~\delta_{lp}~J^{kq} + i~\delta_{kp}~J^{lq} 
+ i~\delta_{lq}~J^{kp} - i~\delta_{kq}~J^{lp};
\label{p2}
\eea
here
$j,k,l = 1,\dots,D-2$
and
$\epsilon = \pm$.

We will need the representation of mass 
$m > 0$ 
and spin $0$ in light-cone coordinates. The Hilbert space is
\be
{\cal H}^{[m,0]} \equiv  {\cal H}^{[m]} 
\equiv L^{2}\left( \R^{D-1}, {d{\bf p}\over 2p^{0}}\right)
\ee
where
$
d{\bf p} \equiv dp^{1}\dots dp^{D-1},
$
$
p^{0} \equiv \sqrt{{\bf p}^{2} + m^{2}}
$
and
$
{\bf p}^{2} = ({\bf p}^{1})^{2} + \cdots ({\bf p}^{D-1})^{2}.
$

One can easily derive the expressions of the Lie algebra generators; they are:
\bea
P^{\mu} = p^{\mu}
\nonumber \\
L^{jk} = - i~\left( p^{j} {\partial \over \partial p^{k}} 
- p^{k} {\partial \over \partial p^{j}} \right)
\nonumber \\
K^{j} \equiv L^{j0} = i~p^{0}~{\partial \over \partial p^{j}}
\eea
where
$j,k = 1,\dots,D-1$.
These operators are defined on
$
{\cal C}^{\infty}(\R^{D-1})
$
where they are essentially self-adjoint; one can easily verify that on this domain the
relations (\ref{p1}) are valid.

Now we perform a change of variables \cite{Di1}:
\be
(p^{1},\dots,p^{D-1}) \rightarrow (p^{+},\tilde{p})
\ee
where
\be
p^{+} \equiv {1\over \sqrt{2}} (p^{0} + p^{D-1}) \qquad 
\tilde{p} = (p^{1},\dots,p^{D-2});
\ee
here, as before, we have denoted
$
p^{0} \equiv \sqrt{{\bf p}^{2} + m^{2}}.
$
We will also denote
$
\tilde{\bf p}^{2} = ({\bf p}^{1})^{2} + \cdots ({\bf p}^{D-2})^{2}.
$

The converse of this change of variables is:
\be
p^{D-1} = {1\over \sqrt{2}} (p^{+} - p^{-}) 
\ee
where we have introduced the notation 
\be
p^{-} = p^{-}(p^{+},\tilde{\bf p}) \equiv { \tilde{\bf p}^{2} + m^{2} \over 2 p^{+}}.
\ee

Now we re-express the measure
$
{d{\bf p}\over 2p^{0}}
$
in the new variables and easily get:
\be
{d{\bf p}\over 2p^{0}} = {d\tilde{p} dp^{+}\over 2p^{+}}.
\label{measure}
\ee

Let us note that for 
$
m > 0
$
the variable
$
p^{+}
$
takes values in the whole positive real axis so, in the new variables the Hilbert space is
\be
{\cal H}^{[m]} 
\equiv L^{2}\left( \R_{+} \times \R^{D-2}, {dp^{+} d\tilde{p} \over 2p^{+}}\right)
\label{h-m}
\ee
where as usual
$
\R_{+} \equiv (0, \infty ).
$
The unitary transformation connecting the two representations is
$
V: L^{2}\left( \R^{D-1}, {d{\bf p}\over 2p^{0}}\right) \rightarrow
L^{2}\left( \R_{+} \times \R^{D-2}, {dp^{+} d\tilde{p} \over 2p^{+}}\right)
$
is:
\be
(Vf)(p^{+},\tilde{p}) = f\left(\tilde{p}, {p^{+} - p^{-} \over \sqrt{2}}\right);
\label{V}
\ee
the inverse of this transformation can be easily provided.

It is a straightforward exercise to derive the expressions of the Lie algebra generators
in the light-cone coordinates:
\bea
P^{j} = p^{j} \qquad
P^{+} = p^{+} \qquad
P^{-} = p^{-} \equiv { \tilde{\bf p}^{2} + m^{2} \over 2 p^{+}}
\nonumber \\
L^{jk} = - i~\left( p^{j} {\partial \over \partial p^{k}} 
- p^{k} {\partial \over \partial p^{j}} \right) \qquad
L^{j+} = i~ p^{+} {\partial \over \partial p^{j}} 
\nonumber \\
L^{j-} = i~\left( p^{-} {\partial \over \partial p^{j}} 
+ p^{j} {\partial \over \partial p^{+}} \right) \qquad
L^{+-} = - i~p^{+}~{\partial \over \partial p^{+}}
\label{lc}
\eea
where
$j,k = 1,\dots,D-2$.
These operators are defined on
$
{\cal D}^{[m]} \equiv {\cal C}^{\infty}(\R_{+} \times \R^{D-2})
$
where they are essentially self-adjoint; one can easily verify that on this domain the
relations (\ref{p2}) are valid.

The extensions of the preceding formulas to the case of arbitrary $m$ (i.e.
$m = 0$ and
$m$ purely imaginary)
can be done using Lemma 3 of \cite{Di2}. We define for any
$
j = 1,\dots,D-1
$
expressions of the type 
$p^{+}$
namely
\be
p^{+}_{j} \equiv {1\over \sqrt{2}} (p^{0} + p^{j})
\ee
and the chart
\be
V_{j} = \{ (p^{+}_{j}, p^{1},\dots,p^{j-1},p^{j+1},\dots,p^{D-1}) 
\in \R_{+} \times \R^{D-2} \}.
\ee

In every such chart there are no singularities and one can consider the corresponding measure defined as in 
(\ref{measure}). The reunion of all these charts is a cover of the whole mass shell
and if we consider a partition of the unity subordinated to this partition then we
can obtain the corresponding measure in light-cone coordinates. 
\section{Quantum Strings and Superstrings}
In this Section we use the preceding construction to give a proper mathematical
definition for the string and superstring systems.
\subsection{The Quantum Bosonic String\label{b-s}}

We introduce the following construction. First we consider the system of Bose oscillators
$
\alpha_{m}^{j}, m \in \Z^{*}, j = 1,\dots,D-2
$
(i.e. we exclude for the moment the value 
$m = 0$). The Hilbert space generated by these operators is
$
{\cal F}^{(\alpha)};
$
we are in the conditions of Subsections \ref{vir-bose} and \ref{vir-multi}. We now consider
the following Hilbert space:
\be
{\cal H}^{(\mu,\alpha)} \equiv {\cal H}^{[\mu]} \otimes {\cal F}^{(\alpha)}
\label{h-b}
\ee
with 
$
{\cal H}^{[\mu]}
$
defined by (\ref{h-m}); we are denoting the mass by 
$
\mu > 0
$ 
to avoid confusion with the indices 
$m \in \Z$.
If we identify the Fock space
$
{\cal F}^{(\alpha)}
$
with its dual (using the scalar product
$
<\cdot,\cdot>_{\alpha}
$) 
then we can describe the preceding Hilbert space 
as the space of Borel maps from
$\R^{3}$
into 
$
{\cal F}^{(\alpha)}
$
square integrable with respect to the scalar product
\be
< f, g> = \int {d{\bf p} \over 2 p^{0}} <f(p), g(p)>_{\alpha}.
\ee 

We define
\be
\alpha_{0}^{j} \equiv p^{j} \quad j = 1,\dots,D-2
\ee
and we can define the Virasoro generators by
\be
L^{(\alpha)}_{m} \equiv {1\over 2}~\sum_{n \in \Z}~:\alpha^{j}_{m-n} \alpha^{j}_{n}: 
- a \delta_{m} \cdot I
\label{l1}
\ee
(i.e. we are in the conditions from Subsection \ref{vir-multi} with
$
\epsilon_{1} = \dots = \epsilon_{D-2} = 1).
$
More rigorously we have
\bea
L^{(\alpha)}_{m} = I_{1} \otimes  
{1\over 2}~\left(\sum_{n \not= 0,m}~:\alpha^{j}_{m-n} \alpha^{j}_{n}: \right)
+ p^{j} \otimes \alpha_{m}^{j} \quad \forall m \not= 0
\nonumber \\
L^{(\alpha)}_{0} = I_{1} \otimes \sum_{n > 0} :\alpha^{j}_{-n} \alpha^{j}_{n}: 
+ {1\over 2} \tilde{\bf p}^{2} \otimes I_{2} - a \cdot I
\label{l0}
\eea
but we an use the more compact definition (\ref{l1}) without risk of confusion. 
By a slight abuse of notation we will called
$
{\cal D}_{0} \simeq {\cal H}^{[\mu]} \otimes {\cal D}_{0}
$
the algebraic Fock space. Then we have
\begin{lemma}
The operators
$
E^{j}: {\cal H}^{(\mu,\alpha)} \rightarrow {\cal H}^{(\mu,\alpha)}, \quad j = 1,\dots,D-2
$
are well defined on the algebraic Fock by the formulas
\be
E^{j} \equiv - i \sum_{m > 0} {1\over m} 
( \alpha_{-m}^{j}~L_{m} - L_{-m} \alpha_{m}^{j})
\label{e}
\ee
and are formally self-adjoint.
\end{lemma}
{\bf Proof:} We have noticed previously that the vectors
$\alpha_{m}^{j}\Psi$
and
$L_{m}\Psi$
(here 
$\Psi \in {\cal D}_{0}$)
are null for sufficiently large 
$m > 0$
(see Proposition \ref{aaP}.) It follows that the sums in (\ref{e}) are in fact 
finite if we consider vectors of the form
$E^{j}\Psi$.
$\qed$

It is convenient to introduce the {\it Hamiltonian} operator according to
\be
H^{(\alpha)} \equiv \sum_{n > 0} :\alpha^{j}_{-n} \alpha^{j}_{n}: =
\sum_{n > 0} n~(a^{j}_{n})^{+} a^{j}_{n}.
\ee

Now we have the main result:
\begin{thm}
Let us define the following operators on
$
{\cal D}_{0}:
$
\bea
P^{\pm} = p^{\pm} \otimes I_{2} \qquad P^{j} = p^{j} \otimes I_{2}
\nonumber \\
J^{kl} = L^{kl} \otimes I_{2} + I_{1} \otimes J^{(\alpha)kl} \qquad
J^{k+} = L^{k+} \otimes I_{2}
\nonumber \\
J^{k-} = L^{k-} \otimes I_{2} + {1\over p^{+}} E^{k} \qquad
J^{+-} = L^{+-} \otimes I_{2};
\label{p-b}
\eea
here
$k,l = 1,\dots,D-2$
as usual and the operators
$
J^{(\alpha)kl}
$
are those defined by the first formula of (\ref{j}). Then these operators are a 
representation of the Poincar\'e Lie algebra iff
$D = 26$
and we consider only the states from the {\bf physical Hilbert space}
\be
{\cal H}^{(\mu,\alpha)}_{phys} \equiv \left\{ \Psi \in {\cal H}^{(\mu,\alpha)} | 
H^{(\alpha)} \Psi = \left( 1 + {\mu^{2} \over 2} \right) \Psi \right\}.
\label{phys-a1}
\ee
\label{bs-inv}
\end{thm}
{\bf Proof:} We have to check the formulas (\ref{p2}). Of course, we will use the
fact the the operators
$
L^{\dots}
$
verify these relations as stated in the previous Section so the non-trivial 
ones must have at least a
$
J^{k-}
$
entry. We are left with the following non-trivial relations to check:
\bea
~[ J^{+-}, J^{k-} ] =  i~J^{k-} \qquad
~[ J^{j+}, J^{k-} ] = - i~J^{jk} - i~\delta_{jk}~J^{+-} 
\nonumber \\
~[ J^{j-}, J^{kl} ] = - i~\delta_{jk}~J^{l-} + i~\delta_{jl}~J^{k-} \qquad
~[ J^{j-}, J^{k-} ] = 0.
\eea
The first three preceding relations can be checked elementary and do not
produce anomalies. Only the last relation is highly non-trivial. To
compute the commutator
$
[ J^{j-}, J^{k-} ] = 0
$
we use the elementary formula
\be
[ AB, CD] = A [ B, C ] D + AC [ B, D ] + [ A, C ] DB + C [ A, D ] B. 
\label{com}
\ee
We need to compute first
\be
[ E^{j}, E^{k} ] = - (C^{jk}_{1} + C^{jk}_{2} - H.c.)
\label{ee}
\ee
where
\be
C^{jk}_{1} = \sum_{m,n > 0} {1\over mn} [ \alpha_{-m}^{j} L_{m}, \alpha_{-n}^{k} L_{n} ] 
\quad
C^{jk}_{2} = \sum_{m,n > 0} {1\over mn} [ \alpha_{-m}^{j} L_{m}, L_{-n} \alpha_{n}^{k} ]
\ee
and we understand that all operators are acting on vectors from the
algebraic Fock space.
These commutators can be computed using the formulas (\ref{com}) and 
(\ref{vir-alpha-D}) with
$
D \rightarrow D - 2.
$
One checks that at
every stage of the computations the sums are in fact finite because we apply the
commutators only on vectors from the algebraic Fock space.
We give only the final results:
\bea
C^{jk}_{1} = \left( \sum_{ m > 0} {m - 1\over 2} \alpha^{j}_{-m} \alpha^{k}_{m}
+ \sum_{ m > n \geq 0} {1\over m} \alpha^{j}_{-m} L_{m-n} \alpha^{k}_{n} \right)
- ( j \leftrightarrow k)
\nonumber \\
C^{jk}_{2} = \sum_{ m \geq n > 0} {1\over m} \alpha^{j}_{-m} L_{m-n} \alpha^{k}_{n}
+ \sum_{ n \geq m > 0} {1\over n} \alpha^{j}_{-m} L_{m-n} \alpha^{k}_{n}
\nonumber \\
+ \sum_{ m > 0} \left[ m {D - 26\over 2} + {1\over m} \left( 2a - {D - 2 \over 12} 
\right) + 1 \right] \alpha^{k}_{-m} \alpha^{j}_{m} + \dots
\eea
where by $\dots$ we mean a term proportional to 
$\delta_{jk}$
which disappears from (\ref{ee}). We finally obtain
\bea
[ E^{j}, E^{k} ] = 
\sum_{ m > 0} \left[ m {D - 26\over 12} + {1\over m} \left( 2a - {D - 2 \over 12} 
\right)  \right] [ \alpha^{j}_{-m} \alpha^{k}_{m} - ( j \leftrightarrow k) ]
\nonumber \\
+ i~(p^{j} E^{k} - p^{k} E^{j}) 
+ 2 \sum_{ m > 0} {1\over m} [\alpha^{j}_{-m} \alpha^{k}_{m} 
- ( j \leftrightarrow k) ] L_{0}.
\label{ee1}
\eea

Let us note that if we use the definition (\ref{l0}) we have: 
$
L_{0} = H^{(\alpha)} + {1\over 2} \tilde{\bf p}^{2} - a 
$
and we observe that the constant $a$ drops out! Now we insert the 
commutator (\ref{ee1}) in the formula
\bea
[ J^{j-}, J^{k-} ] = [ L^{j-}, L^{k-} ] \otimes I_{2} +
{1\over (p^{+})^{2}} [ E^{j}, E^{k} ] 
\nonumber \\
+ \left( L^{j-} {1\over p^{+}} \right) E^{k} - \left( L^{k-} {1\over p^{+}} \right) E^{j}
+  {1\over p^{+}} (L^{j-} E^{k}) - {1\over p^{+}} (L^{k-} E^{j})
\eea
and get the final result
\be
[ J^{j-}, J^{k-} ] = {1\over (p^{+})^{2}} 
\sum_{ m > 0} [ \alpha^{k}_{-m} \alpha^{j}_{m} - ( j \leftrightarrow k) ]
\left[ m {D - 26\over 12} + {1\over m} \left( 2 H^{(\alpha)} - {D - 2 \over 12} 
- {\mu}^{2} \right)  \right];
\label{comm-a1}
\ee
equating to zero we obtain the value of $D$ and also the expression
for the physical Hilbert space. It remains to show that the physical Hilbert
space is Poincar\'e invariant i.e. it is left invariant by the operators 
(\ref{p-b}) from the statement. This follows from
\be
[ H^{(\alpha)}, K^{j} ] = [ L_{0}, K^{j} ] = 0
\ee
and the proof is finished.
$\qed$

Let us remark that the vacuum
$\Omega$
does {\it not} belong to the physical Hilbert space. The preceding system seems to
be closest to what the physical intuition tells us a vibrating string of mass 
$\mu$ 
should be:
the first factor in (\ref{h-b}) describes the translation of the string in 
space-time and the second factor the vibrations of the string in the rest frame.
Because the operator 
$
H^{(\alpha)}
$
has the spectrum
$
\sigma(H^{(\alpha)}) = \Z
$
we obtain that in this case the mass $\mu$ is quantized:
$
\mu^{2} \in 2 \cdot \N.
$
 
However, a different construction is preferred in the literature and is called the
{\it Bosonic string}. Instead of (\ref{h-b}) we take
\be
{\cal H}^{(b)} \equiv 
\left( \oplus_{l \in L}{\cal H}^{[\mu_{l}]} \right)\otimes {\cal F}^{(\alpha)};
\label{h-b1}
\ee
where we the sum is over an unspecified set $L$ (not necessarily finite) of masses. The extensions
of the formulas (\ref{p-b}) to this case are obvious. (In fact,
$
{\cal H}^{(b)}
$
is a direct sum of Hilbert spaces of the type
$
{\cal H}^{(\mu,\alpha)}).
$
The same computation as above
leads to
\be
[ J^{j-}, J^{k-} ] = {1\over (p^{+})^{2}} 
\sum_{ m > 0} [ \alpha^{k}_{-m} \alpha^{j}_{m} - ( j \leftrightarrow k) ]
\left[ m {D - 26\over 12} + {1\over m} \left( 2 H^{(\alpha)} - {D - 2 \over 12} 
- p^{2} \right)  \right]
\label{comm-a2}
\ee
where now
\be
p^{2} = \oplus_{l \in L} \mu^{2}_{l} I_{l}.
\ee

We obtain as before
$D = 26$ 
but the physical Hilbert space is
\be
{\cal H}^{(b)}_{phys} \equiv \left\{ \Psi \in {\cal H}^{(b)} | 
2 (H^{(\alpha)}  - 1)\Psi = p^{2} \Psi \right\}.
\label{phys-a2}
\ee
In this way we get tachyons in the spectrum of the model (for instance the vacuum 
state corresponds to
$p^{2} = -2$).
In this case the expression of
$
J^{kl}
$
from (\ref{p-b}) makes sense only on functions defined in the chart 
$
V_{D-1}
$ 
and of compact support (such that the singularity in
$p^{+} = 0$
is integrable). Similar constructions must be considered in all charts.

We note in the end that the necessity of considering only states lying in the physical 
Hilbert space (\ref{phys-a1}) or (\ref{phys-a2}) appears in the standard literature in a different form e.g. equation (2.3.12) from \cite{GSW}. We also note that the condition 
$
a = 1
$ 
appearing frequently in the literature is not needed. Apparently this fact is known in the literature but we cannot provide an explicit reference on this point.

\subsection{The Neveau-Schwartz Superstring\label{ns-s}}
We generalize the previous arguments for the superstring. In the NS case we consider
the Hilbert space generated by the system of Bose oscillators
$
\alpha_{m}^{j}, m \in \Z^{*}, j = 1,\dots,D-2
$
(i.e. we exclude for the moment the value 
$m = 0$) and the the Fermi oscillators
$
b_{r}^{j}, r \in {1\over 2} + \Z, j = 1,\dots,D-2;
$
The Hilbert space generated by these operators is
$
{\cal F}^{(NS)};
$
we are in the conditions of Subsections \ref{vir-ns} and \ref{vir-multi}. We now consider
the following Hilbert space:
\be
{\cal H}^{(NS)} \equiv {\cal H}^{[\mu]} \otimes {\cal F}^{(NS)}
\label{h-ns}
\ee
with 
$
{\cal H}^{[\mu]}
$
defined by (\ref{h-m}). We define as in the previous Subsection
\be
\alpha_{0}^{j} \equiv p^{j} \quad j = 1,\dots,D-2
\ee
and we are in the conditions of Subsection \ref{vir-multi} so we can define the
Virasoro generators
\be
L^{(\alpha)}_{m} \equiv {1\over 2}~\sum_{n \in \Z}~:\alpha^{j}_{m-n} \alpha^{j}_{n}: 
+ {1\over 2}~\sum_{r \in 1/2 + \Z}~r~:b^{j}_{-r} b^{j}_{m+r}: - a \delta_{m} \cdot I.
\label{l2}
\ee
Then we can define the operators
$
E^{j}, F^{j}: {\cal H}^{(NS)} \rightarrow {\cal H}^{(NS)}, \quad j = 1,\dots,D-2
$
on the algebraic Fock space by the formulas
\be
E^{j} \equiv - i \sum_{m > 0} {1\over m} 
( \alpha_{-m}^{j}~L_{m} - L_{-m} \alpha_{m}^{j})
\label{e1}
\ee
\be
F^{j} \equiv - i \sum_{m \in \Z} K^{jl}_{-m} \alpha_{m}^{l} 
= - i \sum_{m > 0} (\alpha_{-m}^{l} K^{jl}_{m} + K^{jl}_{-m} \alpha_{m}^{l})  
- i K^{jl}_{0} \alpha_{0}^{l}
\label{f1}
\ee
where the operators
$
K^{jl}_{m}
$
have been defined in the Subsection \ref{km}.
We remark that the expression (\ref{e1}) formally coincides with (\ref{e}) but the
expression of
$
L_{m}
$
is in fact different: we have a Fermi contribution in (\ref{l2}).
\be
K^{j} \equiv E^{j} + F^{j}
\ee
and all operators
$
E^{j}, F^{j}, K^{j})
$
are formally self-adjoint. The {\it Hamiltonian} operator also has a Fermi contribution:
\be
H^{(NS)} \equiv \sum_{n > 0} 
:\alpha^{j}_{-n} \alpha^{j}_{n}: + \sum_{r > 0}~r~:b^{j}_{-r} b^{j}_{r}:
= \sum_{n > 0} n~(a^{j}_{n})^{+} a^{j}_{n} + \sum_{r > 0}~r~:b^{j}_{-r} b^{j}_{r}:
\ee

Now we have the main result:
\begin{thm}
Let us define the following operators on
$
{\cal D}^{[m]} \otimes {\cal D}_{0}:
$
\bea
P^{\pm} = p^{\pm} \otimes I_{2} \qquad P^{j} = p^{j} \otimes I_{2}
\nonumber \\
J^{kl} = L^{kl} \otimes I_{2} + I_{1} \otimes J^{(NS)kl} \qquad
J^{k+} = L^{k+} \otimes I_{2}
\nonumber \\
J^{k-} = L^{k-} \otimes I_{2} + {1\over p^{+}} K^{k} \qquad
J^{+-} = L^{+-} \otimes I_{2};
\label{p-ns}
\eea
here
$k,l = 1,\dots,D-2$
as usual and the operators
$
J^{(NS)kl} = J^{(\alpha)kl} + J^{(b)kl}
$
are those defined by the formula (\ref{j}). Then these operators are a 
representation of the Poincar\'e Lie algebra iff
$D = 10$
and we consider only the states from the {\bf physical Hilbert space}
\be
{\cal H}^{(NS)}_{phys} \equiv \left\{ \Psi \in {\cal H}^{(NS)} | 
H^{(NS)} \Psi = {1\over 2} ( 1 + \mu^{2}) \Psi \right\}.
\ee
\end{thm}
{\bf Proof:} As in the previous Subsection we check the formulas (\ref{p2}) and the 
obstructions can come only from the commutator
$
[ J^{j-}, J^{k-} ].
$
The commutator
$
[ E^{j}, E^{k} ]
$
can be obtained from the corresponding formula of the preceding Subsection with the
substitution
$
{D-2 \over 12} \rightarrow {D-2 \over 8} 
$
in the commutators of
$L_{m}$'s
as it follows by comparing (\ref{vir-alpha-D}) to (\ref{lll}). We get in this way easily:
\bea
[ E^{j}, E^{k} ] = 
\sum_{ m > 0} \left[ m {D - 10\over 8} + {1\over m} \left( 2a - {D - 2 \over 8} 
\right)  \right] \alpha^{k}_{-m} \alpha^{j}_{m} - ( j \leftrightarrow k)
\nonumber \\
+ i~(p^{j} E^{k} - p^{k} E^{j}) 
+ 2 \sum_{ m > 0} {1\over m} [ \alpha^{j}_{-m} \alpha^{k}_{m} 
- ( j \leftrightarrow k) ] L_{0}.
\eea

To obtain the expression
$
[ K^{j}, K^{k} ]
$
we still have to compute the commutators
$
[ F^{j}, F^{k} ] 
$
and
$
[ E^{j}, F^{k} ]
$
for which we use again (\ref{com}). After a tedious but straightforward algebra we get
\bea
[ J^{j-}, J^{k-} ] = {1\over (p^{+})^{2}} 
\sum_{ m > 0} [\alpha^{k}_{-m} \alpha^{j}_{m} - ( j \leftrightarrow k) ]
\left[ m {D - 10\over 8} + {1\over m} \left( 2 H^{(NS)} - {D - 2 \over 8} 
- \mu^{2} \right)  \right] 
\nonumber \\
+ {D - 10\over 4} \sum_{ r > 0} r (2r -1) 
\left[ b^{j}_{-r} b^{k}_{r} - ( j \leftrightarrow k) \right]
+ {1\over p^{+}}~(2 H^{(NS)} - 1 - \mu^{2})~K^{jk}_{0};
\eea
to obtain this formula we use in an essential way the formula (\ref{kkl-ns}) of Section \ref{km}. Now we equate to zero the right hand side and the theorem follows.
$\qed$ 

The vacuum
$\Omega$
does {\it not} belong to the physical Hilbert space in this case also. As in the
preceding Subsection, a different construction is preferred in the literature and 
is called the {\it Neveau-Schwartz superstring}. Instead of (\ref{h-ns}) we take
\be
{\cal H}^{(NS)} \equiv 
\left( \oplus_{l \in L}{\cal H}^{[\mu_{l}]} \right)\otimes {\cal F}^{(NS)};
\label{h-ns1}
\ee
where we the sum is over an unspecified set $L$ of masses. The extensions
of the formulas (\ref{p-ns}) to this case are obvious. The same computation as above
leads to
\bea
[ J^{j-}, J^{k-} ] = {1\over (p^{+})^{2}} 
\sum_{ m > 0} [\alpha^{k}_{-m} \alpha^{j}_{m} - ( j \leftrightarrow k) ]
\left[ m {D - 10\over 8} + {1\over m} \left( 2 H^{(NS)} - {D - 2 \over 8} 
- p^{2} \right)  \right] 
\nonumber \\
+ {D - 10\over 4} \sum_{ r > 0} r (2r -1) 
\left[ b^{j}_{-r} b^{k}_{r} - ( j \leftrightarrow k) \right]
+ {1\over p^{+}}~(2 H^{(NS)} -1 - p^{2})~K^{jk}_{0};
\eea
where now
\be
p^{2} = \oplus_{l \in L} \mu^{2}_{l} I_{l}.
\ee

We obtain as before
$D = 10$ 
but the physical Hilbert space is \cite{GSW}
\be
{\cal H}^{(NS)}_{phys} \equiv \left\{ \Psi \in {\cal H}^{(NS)} | 
(2 H^{(NS)}  - 1)\Psi = p^{2} \Psi \right\}.
\ee
In this way we get tachyons in the spectrum of the model (for instance the vacuum state
corresponds to
$p^{2} = -1$).
One can eliminate the tachyons imposing the GSO condition \cite{GSW} namely, considering that
the physical Hilbert space is the subspace of
$
{\cal H}^{(NS)}_{phys}
$
generated by odd numbers of $b$ oscillators and arbitrary numbers of 
$\alpha$ 
oscillators; this subspace is again Poincar\'e invariant.
The parameter $a$ remains unconstrained in this case also.
\subsection{The Ramond Superstring\label{r-s}}
In the Ramond case we consider the Hilbert space generated by the system of Bose oscillators
$
\alpha_{m}^{j}, m \in \Z^{*}, j = 1,\dots,D-2
$
(i.e. we exclude for the moment the value 
$m = 0$) and the  the Fermi oscillators
$
d_{m}^{j}, m \in \Z, j = 1,\dots,D-2;
$
the Hilbert space generated by these operators is
$
{\cal F}^{(R)};
$
we are in the conditions of Subsections \ref{vir-ns} and \ref{vir-multi}. We consider
the following Hilbert space:
\be
{\cal H}^{(R)} \equiv {\cal H}^{[\mu]} \otimes {\cal F}^{(R)}
\label{h-r}
\ee
with 
$
{\cal H}^{[\mu]}
$
defined by (\ref{h-m}). We define
\be
\alpha_{0}^{j} \equiv p^{j} \quad j = 1,\dots,D-2
\ee
and we are in the conditions of Subsection \ref{vir-multi} so we can define the
Virasoro generators
\be
L^{(\alpha)}_{m} \equiv {1\over 2}~\sum_{n \in \Z}~:\alpha^{j}_{m-n} \alpha^{j}_{n}: 
+ {1\over 2}~\sum_{n \in \Z}~n~:d^{j}_{-n} d^{j}_{m+n}: 
+ \left( {D\over 16} - a \right)~\delta_{m} \cdot I;
\label{l3}
\ee
we remark that we have included the shift (\ref{shift}) of 
$L_{0}$
such that we have the canonical form for the $2$-cocycle of the Virasoro algebra.
Then we can define the operators
$
E^{j}, F^{j}, K^{j}: {\cal H}^{(NS)} \rightarrow {\cal H}^{(NS)}, \quad j = 1,\dots,D-2
$
on the algebraic Fock by the same formulas as in the preceding Subsection 
(however the Virasoro operators are different).

In this case the {\it Hamiltonian} operator also has a Fermi contribution:
\be
H^{(R)} \equiv \sum_{n > 0} 
:\alpha^{j}_{-n} \alpha^{j}_{n}: + \sum_{n > 0}~n~:d^{j}_{-n} d^{j}_{n}:
= \sum_{n > 0} n~(a^{j}_{n})^{+} a^{j}_{n} + \sum_{n > 0}~n~:d^{j}_{-n} d^{j}_{n}:
\ee

Now the main result is:
\begin{thm}
Let us define the following operators on
$
{\cal D}^{[m]} \otimes {\cal D}_{0}:
$
\bea
P^{\pm} = p^{\pm} \otimes I_{2} \qquad P^{j} = p^{j} \otimes I_{2}
\nonumber \\
J^{kl} = L^{kl} \otimes I_{2} + I_{1} \otimes J^{(R)kl} \qquad
J^{k+} = L^{k+} \otimes I_{2}
\nonumber \\
J^{k-} = L^{k-} \otimes I_{2} + {1\over p^{+}} K^{k} \qquad
J^{+-} = L^{+-} \otimes I_{2};
\label{p-r}
\eea
here
$k,l = 1,\dots,D-2$
as usual and the operators
$
J^{(R)kl} = J^{(\alpha)kl} + J^{(d)kl}
$
are those defined by the formula (\ref{j}). Then these operators are a 
representation of the Poincar\'e Lie algebra iff
$D = 10$
and we consider only the states from the {\bf physical Hilbert space}
\be
{\cal H}^{(R)}_{phys} \equiv \left\{ \Psi \in {\cal H}^{(NS)} | 
H^{(NS)} \Psi = {\mu^{2}\over 2} \Psi \right\}.
\ee
\end{thm}
{\bf Proof:} Formally, the content of this theorem coincides with the previous theorem.
Similar computations, making use of (\ref{kkl-r}) this time, lead to:
\bea
[ J^{j-}, J^{k-} ] = {1\over (p^{+})^{2}} 
\sum_{ m > 0} [ \alpha^{k}_{-m} \alpha^{j}_{m} - ( j \leftrightarrow k) ]
\left[ m {D - 10\over 8} + {1\over m} \left( 2H^{(R)} - \mu^{2} \right)  \right]
\nonumber \\
+ {D - 10\over 2} \sum_{ n > 0} n^{2} 
\left[ d^{j}_{-n} d^{k}_{n} - ( j \leftrightarrow k) \right]
+ {1\over p^{+}}~\left(2 H^{(R)} - \mu^{2} \right)~K^{jk}_{0}
\eea
and equating to zero the right hand side and the theorem follows.
$\qed$ 

In the Ramond model the vacuum 
$\Omega$
belongs to the physical Hilbert space. As in the
preceding Subsection, a different construction is preferred in the literature and 
is called the {\it Ramond superstring}. Instead of (\ref{h-r}) we take
\be
{\cal H}^{(R)} \equiv 
\left( \oplus_{l \in L}{\cal H}^{[\mu_{l}]} \right)\otimes {\cal F}^{(R)};
\label{h-r1}
\ee
where we the sum is over an unspecified set $L$ of masses. The extensions
of the formulas (\ref{p-r}) to this case are obvious. We obtain as before
$D = 10$ 
but the physical Hilbert space is
\be
{\cal H}^{(R)}_{phys} \equiv \left\{ \Psi \in {\cal H}^{(R)} | 
H^{(R)} \Psi = {p^{2}\over 2} \Psi \right\}.
\ee
In this way we do not get tachyons in the spectrum of the model.
 
\subsection{Other Superstring Models}
From the preceding two Subsections it is clear that formulas of the type
(\ref{kkl-ns}) and (\ref{kkl-r}) are essential for establishing Lorentz
invariance. We investigate now if such formulas can be valid for more
general cases. More precisely, suppose that we have a $N$-dimensional
representation 
$
\sigma_{jk}, j,k = 1,\dots, D
$
of the algebra
$
so(D)
$
such that
\be
\sigma_{jk}^{T} = - \sigma_{jk};
\ee
then we can define the associated Kac-Moody algebras 
$
K^{jk}_{m}(\sigma)
$
according to the formulas (\ref{km-b}) and (\ref{km-d}) respectively. 
We are interested if in some special cases formulas of the type
(\ref{kkl-ns}) and (\ref{kkl-r}) hold.  A necessary condition is that
the terms quadratic in the operators $b$ (resp. $d$) cancel identically.
It is not very difficult to prove that this requirement is equivalent to
\be
\sum_{r_{1} + \dots + r_{4} = 0} X^{jk}_{p_{1}r_{1};\dots;p_{4}r_{4}} 
:b^{p_{1}}_{r_{1}} \dots b^{p_{4}}_{r_{4}}: = 0
\label{xb}
\ee
where
\be
X^{jk}_{p_{1}r_{1};\dots;p_{4}r_{4}} \equiv
(r_{1} + r_{2}) (\sigma_{jl})_{p_{1}p_{2}} (\sigma_{kl})_{p_{3}p_{4}} 
+ (r_{3} - r_{4}) (\sigma_{jk})_{p_{1}p_{2}}~\delta_{p_{3}p_{4}}
\label{x}
\ee
in the Neveau-Schwartz case and similar relations for the Ramond case.
The relation (\ref{xb}) is equivalent to
\be
X^{jk}_{p_{1}r_{1};\dots;p_{4}r_{4}} - (1 \leftrightarrow 3) 
- (1 \leftrightarrow 4) - (2 \leftrightarrow 3) - (2 \leftrightarrow 4) 
+ (1 \leftrightarrow 3, 2 \leftrightarrow 4) = 0. 
\ee
One inserts here the definition (\ref{x}) and eliminates 
$
r_{4} = - (r_{1} + r_{2} + r_{3});  
$
the result is an equation of the form
\be
r_{1} E^{(1)jk}_{p_{1}\dots;p_{4}} + r_{2} E^{(2)jk}_{p_{1}\dots;p_{4}} 
+ r_{3} E^{(3)jk}_{p_{1}\dots;p_{4}} = 0
\ee
so we obtain three relations
\be
E^{(a)jk}_{p_{1}\dots;p_{4}} = 0 \quad a = 1,2,3.
\label{eee}
\ee

One can easily see that the relation
\be
E^{(1)} + E^{(2)} - E^{(3)} = 0
\ee
is equivalent to
\be
(\sigma_{jl})_{ab} (\sigma_{kl})_{cd}  - (j \leftrightarrow k) = 
\delta_{bd}  (\sigma_{jk})_{ac} - (a \leftrightarrow b) 
- (c \leftrightarrow d) + (a \leftrightarrow b, c \leftrightarrow d)
\label{repr}
\ee
and conversely (\ref{repr}) is implies (\ref{eee}). Moreover, if we have
(\ref{repr}) the one can prove that relations of the type (\ref{kkl-ns})
and (\ref{kkl-r}) hold and we have Lorentz invariance theorems like in
the preceding Subsection in $10$ dimensions. So the key relation
(\ref{repr}) must be analyzed in the case 
$D = 8$.
We note that in this case one should
modify in an appropriate way the expression (\ref{j}) for the generators
of the rotations
$J^{(b)jk}$
and
$J^{(d)jk}$.

One can obtain an important consequence of (\ref{repr}) if we take 
$b = c$
and sum over 
$b = 1, \dots, N$. 
We obtain 
\be
[ \sigma_{jl}, \sigma_{kl} ] = (2 - N) \sigma_{jk};
\ee
on the other hand we have from the representation property of 
$\sigma_{jk}$
\be
[ \sigma_{jl}, \sigma_{kl} ] = (2 - D) \sigma_{jk}
\ee
so we conclude that the representation  
$
\sigma_{jk}
$ 
should be $D$-dimensional, i.e. we need to consider only the
representations of dimension $8$ of the algebra 
$so(8)$. 
It is known that there are four (non-equivalent) such representations:
the vector representation (which we have already used in the preceding
Subsection), the adjoint representation and two the spinor
representations 
${\bf 8}_{s}$ 
and 
${\bf 8}_{c}$ 
of opposite chirality. It seems that the identity (\ref{repr}) is valid
for the for the spinor representations also but the details are not
easily found in the literature so we provide an elementary analysis.
First, it is clear that if we multiply (\ref{repr}) with
$
M_{dc}
$
and make the summation we obtain an equivalent relation
\be
Tr(\sigma_{jl} M) \sigma_{kl} - (j \leftrightarrow k) = [ M - M^{T}, \sigma_{jk} ], \quad \forall M.
\ee

If $M$ is symmetric then the preceding relation is an identity. So (\ref{repr}) is equivalent to
\be
Tr(\sigma_{jl} M) \sigma_{kl} - (j \leftrightarrow k) = 2~[ M, \sigma_{jk} ]
\label{m}
\ee
for all antisymmetric 
$N \times N$-
matrices $M$. Now the number of 
$N \times N$-
antisymmetric matrices is
$
{N(N-1) \over 2};
$
on the other hand the number of matrices
$
\sigma_{jk}
$
is
$
{D(D-1) \over 2}.
$
But we have already established that
$N = D$
so if the matrices
$
\sigma_{jk}
$
are linear independent the relation (\ref{m}) is equivalent to
\be
Tr(\sigma_{jl} \sigma_{pq}) \sigma_{kl} - (j \leftrightarrow k) = 2~[ \sigma_{pq}, \sigma_{jk} ], 
\quad \forall p,q.
\label{m1}
\ee

In particular is is easy to see that the fundamental representation
$E_{jk}$
verifies the preceding identity. We check the identity for the spinor representations.
According to \cite{GSW} one can describe the spinor representations of the algebra
$
so(2n)
$
considering the Fermi Fock space $S$ generated by the operators
$
b_{j}, b_{j}^{*}, j = 1,\dots,n;
$
we have the CAR algebra:
\be
\{ b_{j}, b_{k} \} = 0 \quad \{ b_{j}, b^{*}_{k} \} = \delta_{jk}.
\ee

Next, we define the operators
\bea
\gamma_{j} = b_{j} + b_{j}^{*}, \quad j = 1, \dots, n
\nonumber \\
\gamma_{j} = -i~(b_{n-j} - b_{n-j}^{*}), \quad j = n+1, \dots, 2n
\eea
and prove immediately that they form a 
$2^{n}$-
dimensional representation of the Clifford algebra
$C(2n,0)$
i.e. we have
\be
\{ \gamma_{j}, \gamma_{k} \} = \delta_{jk} \cdot I.
\ee

Then a representation of the algebra
$so(2n)$
is given by the operators
\be
\sigma_{jk} = {1\over 4} [ \gamma_{j}, \gamma_{k} ].
\ee

This representation is not irreducible. In fact let us denote by
$
S_{+}
$
(resp.
$
S_{-})
$
the subspaces of $S$ generated by applying an even (resp. odd) number of creation operators
$b_{j}^{*}$
on the vacuum. The projectors on these subspaces will be denoted by
$P_{\pm}$.
It is easy to see that these two subspaces are left invariant by the representation
$
\sigma_{jk}
$
so it makes sense to define the restrictions
$
\sigma^{\pm}_{jk}.
$
The operators
$
\sigma^{\pm}_{jk}
$
are immediately seen to be linear independent. It is also easy to prove that
$
dim(S_{+}) = dim(S_{-}) = 2^{n-1}
$
i.e. both representations
$
\sigma^{\pm}_{jk}.
$
are of the same dimension
$2^{n-1}$;
these are, by definition, the spinor representations of the algebra
$so(2n)$.
Finally we prove:
\be
Tr(\sigma^{\pm}_{jk} \sigma_{pq}^{\pm}) = 2^{n-3}~(\delta_{kp} \delta_{jq} - \delta_{jp} \delta_{kq}).
\label{lamb}
\ee

Indeed, because the left hand side is a 
$SO(2n)$-
invariant tensor and because of the antisymmetry properties we know that the right hand side
must have the form
$
\lambda~(\delta_{kp} \delta_{jq} - \delta_{jp} \delta_{kq});
$
to determine the constant
$\lambda$
we consider a particular case, say
$
p = j, q = k \quad j \not= k
$
and we obtain
\be
\lambda = - Tr(\sigma_{jk}^{2} P_{\pm}) = {1\over 4} Tr(P_{\pm})
= {1\over 4} dim(S_{\pm}) = 2^{n-3}.
\ee

It follows that only for 
$n = 4$
we have
$\lambda = 2$
and in this case if we use (\ref{lamb}) in (\ref{m1}) we obtain an identity. It follows 
that the two spinor representations of
$so(8)$
verify the identity (\ref{m}) so they can be used to construct supersymmetric string
models as in the preceding two Subsections. These models are considered more consistent
because we quantize using Fermi statistics oscillators pertaining to spinor representations
so we are in agreement with spin-statistics correspondence. Moreover these models 
exhibit also supersymmetry invariance.

There is yet another possibility of constructing consistent models, namely by modifying
the Bosonic string from Subsection  \ref{b-s}. We consider that we have another 
representation of the Virasoro algebra 
$
L_{m}^{c}
$
with central charge $c$ acting in the Hilbert space
$
{\cal H}^{c};
$
we consider the Hilbert space 
$
{\cal H}^{\mu,\alpha} \otimes {\cal H}^{c}
$ 
where 
$
{\cal H}^{\mu,\alpha}
$ 
is given by (\ref{h-b}) and modify the Virasoro algebra given by (\ref{l1}) according to
$
L_{m} \rightarrow L_{m} + L_{m}^{c}.
$
Because the central charges are additive the new Virasoro algebra will have the 
central charge 
$D - 2 + c$
so the consistency condition is in this case 
\be
c = 26 - D
\ee
and the expression of the physical Hilbert space from theorem \ref{bs-inv} remains the same.
In particular if we want a model in
$D = 10$
dimensions we must find out a representation of the Virasoro algebra with central charge
$c = 16$.
It is known that such representations can be found for the groups
$SO(32)$
and
$E_{8} \times E_{8}$
using Sugawara construction \cite{Ot}. This new possibility is used in the construction of the heterotic string models.

For the description of closed strings a doubling of the Bose oscillators
$
\alpha_{m}^{j}
$
appears corresponding to the left and right oscillator modes. Composing in various ways the models described above one can obtain the well-known string models of type I, IIA, IIB and heterotic.

\section{Covariant Quantization of Strings and Superstrings}

One can construct a manifestly covariant formalism also \cite{GSW}. The idea 
is to take in Subsection \ref{vir-multi} the case
$
\epsilon_{0} = - 1, \epsilon_{1} = \dots = \epsilon_{D-1} = 1.
$
In this way the Hilbert space will have states of negative or zero norm. 
So we consider that we have the family of operators:
$
\alpha^{\mu}_{m}, m \in \Z^{*}, \mu = 0,\dots,D
$
acting in the Hilbert space
$
{\cal F}^{(\alpha)}_{\rm cov}
$
such that:
\bea
[ \alpha^{\mu}_{m}, \alpha^{\nu}_{n} ] = - \eta_{\mu\nu}~m~\delta_{m+n} \cdot I, 
\quad \forall m, n
\nonumber \\
\alpha^{\mu}_{m} \Omega = 0, \quad m > 0
\nonumber \\
(\alpha^{\mu}_{m})^{+} = \alpha^{\mu}_{-m} \quad \forall m;
\eea
this Hilbert space will not be positively defined. Define the Virasoro operators
\be
\bar{L}_{m} \equiv - {1\over 2}~\eta_{\mu\nu}~\sum_{n \not= 0,m}~
:\alpha^{\mu}_{m-n} \alpha^{\nu}_{n}: 
\label{vir-bar}
\ee
and we have the following commutation relations:
\be
[ \bar{L}_{m}, \bar{L}_{n} ] = (m - n) \bar{L}^{m+n} 
+ D~{ m (m^{2} - 1) \over 12}~\delta_{m+n}~\cdot I. 
\ee
\begin{prop}
We consider
$
k \in \R^{D}
$
and recursively define the operators
$
U_{n}(k), n \in \N
$
according to
\be
U_{0} = I \qquad
U_{n}(k) = {1 \over n} \sum_{l=1}^{n} U_{n-l}(k) k\cdot \alpha_{l}.
\ee
For convenience we define 
$
U_{n} = 0 \quad \forall n < 0.
$
Then the following relation is valid:
\be
[ \alpha^{\mu}_{m}, U_{n}(k) ] = \theta(m - 1)~k^{\mu}~U_{m+n}(k)
\ee
where
$
\theta(m) 
$
is the usual Heaviside function. 
\end{prop}
The proof is easily done by induction on $n$. Let us remark that the expressions
$
U_{n}(k)
$ 
are the coefficients of the formal series in
$
z \in \C:
$
\be
U(z,k) \equiv e^{A(z,k)}  \qquad
A(z,k) \equiv \sum_{n \geq 1} {1\over n} k\cdot \alpha_{n} z^{n}.
\ee

The recurrence relation from the statement of the proposition can be found if we compute the derivative of 
\be
U(z,k) = \sum_{n \geq 1} U_{n}(k) z^{n}
\ee
in two ways. The explicit relation is
\be
U_{n}(k) = \sum_{p \geq 0} {1\over p!} \sum_{i_{1},\dots,i_{p} > 0}
\sum_{i_{1}+\cdots i_{p}=n}~{1 \over i_{1} \dots i_{p}} 
(k\cdot \alpha_{i_{1}}) \cdots (k\cdot \alpha_{i_{p}})
\ee
but it is convenient to work with the recurrence relation and not with the explicit expression given above.

The operators
$
U_{n}(k)
$
are leaving the algebraic Fock space 
$
{\cal D}_{0}
$
invariant and moreover for every
$
\Psi \in {\cal D}_{0}
$
we have 
\be
U_{n}(k) \Psi = 0
\ee
for sufficiently large $n$. It is useful to expressed the formal series relations:
\bea
[ U(z,k)^{\dagger}, U(z,k^{\prime}) ] = 0 \quad {\rm for} \quad k\cdot k^{\prime} = 0 
\nonumber \\
U(z,k) U(z,k^{\prime}) = U(z,k+k^{\prime}). 
\eea
in terms of the 
$
U_{p}(k)
$
operators. In particular we have
\be
\sum_{p \in \Z} U_{n-p}(k)~U_{p}(k^{\prime}) = U_{n}(k+k^{\prime}).  
\ee
Less elementary are the following sum relations:
\bea
\sum_{p \in \Z} p~U_{n-p}(k)~U_{m+p}(k^{\prime}) 
= \sum_{l > 0} U_{m+n-l}(k+k^{\prime})~k^{\prime}\cdot\alpha_{l} - m~U_{m+n}(k+k^{\prime})  
\nonumber \\
\sum_{p \in \Z} p~U_{n-p}(k)~U_{m+p}(-k) 
= -\theta(m+n) k\cdot\alpha_{m+n} - m~\delta_{m+n} \cdot I
\nonumber \\
\sum_{p \in \Z} p~U_{n-p}(\alpha k)~U_{m+p}(\beta k^{\prime}) 
= {n\beta - m \alpha \over \alpha + \beta}~U_{m+n}((\alpha + \beta) k)
\qquad \alpha, \beta \in \R, \alpha + \beta \not= 0.
\label{sum-u}
\eea 

Now we have
\begin{prop}
The operators
$
V_{n}(k), n \in \Z 
$
are well defined on the algebraic Fock space according to the relations
\be
V_{n}(k) = \sum_{p \in \Z} U_{p-n}(-k)^{\dagger}~U_{p}(k). 
\ee
\end{prop}
Indeed the sum over $p$ is in fact finite because we have
$
U_{n}(k) \Psi = 0
$
for sufficiently high $n$ if 
$
\Psi \in {\cal D}_{0}.
$

The expressions
$
V_{n}(k)
$ 
are the coefficients of the formal series
\be
V(z,k) \equiv U(z,-k)^{\dagger}~U(z,k) 
\ee

We have analogue elementary properties: the operators
$
V_{n}(k)
$
are leaving the algebraic Fock space 
$
{\cal D}_{0}
$
invariant and moreover for every
$
\Psi \in {\cal D}_{0}
$
we have 
\be
V_{n}(k) \Psi = 0
\ee
for sufficiently large $n$. We have
\bea
V_{n}(k)^{\dagger} = V_{-n}(-k).
\nonumber \\
~[ \alpha^{\mu}_{m}, V_{n}(k) ] = (1 - \delta_{m})~k^{\mu}~V_{m+n}(k).
\nonumber \\
\sum_{p \in \Z} V_{n-p}(k)~V_{p}(k^{\prime}) = V_{n}(k+k^{\prime}).  
\eea

Sum relations of the type (\ref{sum-u}) can be found for the 
$
V_{p}(k)
$ 
operators.

Now we can derive the conformal properties of these operators i.e. the commutation relations with the Virasoro operators.
\begin{prop}
The following relation is true:
\be 
[ \bar{L}_{m}, V_{n}(k) ] = - (c_{m} k^{2} + m + n)~V_{m+n}
\ee
where we have defined
\be
c_{m} \equiv \cases{ {m - 1\over 2}, & for m $>$ 0 \cr
{m + 1\over 2}, & for m $<$ 0 \cr
0, & for m = 0 \cr}.
\ee 

In particular if
$
k^{2} = 0
$
we have
\be 
[ \bar{L}_{m}, V_{n}(k) ] = - (m + n)~V_{m+n}
\ee
i.e. the operators 
$
V_{n}(k)
$
have conformal dimension $0$.
\end{prop}
The computation is straightforward: we first compute the commutations relation of the Virasoro operators with 
$
U_{n}(k)
$
and then we use the definition of the operators
$
V_{n}(k).
$
We only note a discrepancy with the standard literature where it is asserted that these operators have defined conformal dimension for any
$
k \in \R^{D};
$
the origin of this discrepancy can be traced to the coefficient 
$
c_{m}
$ 
which is different from the standard literature. Fortunately, only the case
$
k^{2} = 0
$
is needed for the construction of the DDF operators.

We are approaching the definition of the DDF operators. First we define the operators
\be
\bar{V}^{\mu}_{n}(k) \equiv \sum_{p > 0} 
~[ \alpha^{\mu}_{-p} V_{n+p}(k) + V_{n-p}(k) \alpha^{\mu}_{p} ]
\ee
and we have 
\begin{prop}
Let
$
k \in \R^{D}, \quad k^{2} = 0.
$
Let us define the following operators:
\be
\bar{A}^{\mu}_{m} \equiv \bar{V}^{\mu}_{m}(mk).
\ee
Then the following relations are verified:
\bea
[ \bar{A}^{\mu}_{m}, \bar{A}^{\nu}_{n} ] = - \eta^{\mu\nu}~m~\delta_{m+n}\cdot I
+ k^{\mu}~\bar{V}^{\nu}_{m,n}(k) - k^{\nu}~\bar{V}^{\mu}_{n,m}(k)
\nonumber \\
~[ \bar{L}_{m}, \bar{A}^{\mu}_{n} ] = - n~\bar{V}^{\mu}_{m+n}
+ {m(m-1)\over 2} k^{\mu}~V_{m+n}(nk)
\nonumber \\
\bar{A}_{n}^{\mu}(k)^{\dagger} = \bar{A}^{\mu}_{-n}(-k)
\nonumber \\
\bar{A}^{\mu}_{n} \Omega = 0 \quad \forall m > 0
\nonumber \\
\bar{A}^{\mu}_{0} = 0
\eea
where the explicit expressions 
$
\bar{V}^{\nu}_{m,n}(k)
$
are not important.
\end{prop}
To construct the DDF operators we have to include the kinematic degrees of freedom also.
We define the Hilbert space
$
{\cal H}^{[\mu,\alpha]}_{\rm cov} \equiv 
{\cal H}^{[\mu]} \otimes {\cal F}^{(\alpha)}_{\rm cov}
$
where 
$
{\cal H}^{[\mu]}
$
is the Hilbert space of a particle of mass 
$
\mu
$
and spin $0$ and 
$
{\cal F}^{(\alpha)}_{\rm cov}
$
is the Fock space defined at the beginning of this Section. We use the convention
\be
\alpha^{\mu}_{0} = p^{\mu}
\ee
and define the covariant Virasoro operators
\be
L^{(\alpha)}_{m} \equiv - {1\over 2}~\eta_{\mu\nu}~\sum_{n \in \Z}~
:\alpha^{\mu}_{m-n} \alpha^{\nu}_{n}: - a\delta_{m} \cdot I
\label{vir-cov}
\ee
such that we have the following commutation relations:
\be
[ L^{(\alpha)}_{m}, L^{(\alpha)}_{n} ] = (m - n) L^{(\alpha)}_{m+n} 
+ D~{ m (m^{2} - 1) \over 12}~\delta_{m+n}~\cdot I. 
\ee
In this Hilbert space we have a natural action of the Poincar\'e
algebra. This Hilbert space will have states of negative or zero norm.
Now we can define the DDF operators:
\begin{thm}
Let
$
k \in \R^{D}
$
be such that
$
k^{2} = 0.
$
Let us define in 
$
{\cal H}^{[\mu,\alpha]}_{\rm cov}
$
the operators
\be
V^{\mu}_{n}(k) \equiv \bar{V}^{\mu}_{n}(k) + p^{\mu}~V_{n}(k)
\ee
and
\be
A^{\mu}_{n} \equiv V^{\mu}_{n}(nk)
\ee
Then the following relations are true:
\bea
[ A^{\mu}_{m}, A^{\nu}_{n} ] = - \eta^{\mu\nu}~m~\delta_{m+n}\cdot I
+ k^{\mu}~V^{\nu}_{m,n}(k) - k^{\nu}~V^{\mu}_{n,m}(k)
\nonumber \\
~[ L^{(\alpha)}_{m}, A^{\mu}_{n} ] = - n~(1 + k\cdot p)~V^{\mu}_{m+n}
+ {m(m-1)\over 2}~n~k^{\mu}~V_{m+n}(nk)
\eea
where the explicit expressions 
$
V^{\nu}_{m,n}(k)
$
are not important. In particular consider that
$
k \in \R^{D}
$
also verifies
$
k^{j} = 0, \quad j = 1,\cdots,D-1, \quad k\cdot p = - 1
$
(e.g.
$
k^{+} = 0, \quad k^{-} = - {1\over p^{+}}, \quad k^{j} = 0, \quad j = 1,\cdots,D-1
$)
then the operators
$
A^{j}_{n}, \quad j = 1,\cdots,D-1
$
verify
\bea
[ A^{j}_{m}, A^{k}_{n} ] = \delta_{jk}~m~\delta_{m+n}\cdot I
\nonumber \\
~[ L^{(\alpha)}_{m}, A^{j}_{n} ] = 0, \quad m \not= 0 \qquad
[ L^{(\alpha)}_{0}, A^{j}_{n} ] = - n~A^{j}_{n}
\nonumber \\
(A_{n}^{j})^{\dagger} = A^{j}_{-n}
\nonumber \\
A^{j}_{n} \Omega = 0 \quad \forall m > 0
\nonumber \\
A^{j}_{0} = p^{j}
\label{A-b}
\eea
so they verify the same algebra as the operators
$
\alpha^{j}_{m}.
$
\label{DDF-b}
\end{thm}

The {\it DDF operators}
$
A^{j}_{n}
$
are the $z$-independent component of the vertex operator
\be 
\dot{X}^{j}(z)~e^{ik\cdot X(z,nk)}
\ee
where
\be
X^{\mu}(z) \equiv \sum_{n \not= 0} {1\over n} \alpha^{\mu}_{n}~z^{n} + p^{\mu}~ln(z)
\qquad
\dot{X}^{j}_{n} \equiv \sum_{n \not= 0} \alpha^{j}_{n}~z^{n} + p^{j}.
\ee
\section{The Covariant Quantum Bosonic String\label{b-s-cov}}

We describe the Bosonic string (see Section \ref{b-s}) using the Hilbert space bundle formalism \cite{Va}. First we consider the system of Bose oscillators
$
\alpha_{m}^{j}, m \in \Z^{*}, j = 1,\dots,D-2
$
(i.e. we exclude for the moment the value 
$m = 0$). The Hilbert space generated by these operators is
$
{\cal F}^{(\alpha)};
$
we are in the conditions of Subsections \ref{vir-multi} and \ref{b-s}. We now consider
the following Hilbert space:
\be
{\cal H}^{(\mu,\alpha)} \equiv {\cal H}^{[\mu]} \otimes {\cal F}^{(\alpha)}
\label{h-b-cov}
\ee
with 
$
{\cal H}^{[\mu]}
$
defined as above. We define
\be
\alpha_{0}^{j} \equiv p^{j} \quad j = 1,\dots,D-2
\ee
and the {\it transversal Virasoro generators} by
\be
L^{T}_{m} \equiv {1\over 2}~\sum_{n \in \Z}~:\alpha^{j}_{m-n} \alpha^{j}_{n}: 
- a \delta_{m} \cdot I
\label{l1-trans}
\ee
which verify the Virasoro algebra with central charge
$
D - 2.
$
Then we define similarly to (\ref{e}) the operators
$
E^{j}(p): {\cal H}^{(\mu,\alpha)} \rightarrow {\cal H}^{(\mu,\alpha)}, \quad j = 1,\dots,D-2
$
on the algebraic Fock according to the formulas
\be
E^{j}(p) \equiv - i \sum_{m > 0} {1\over m} 
( \alpha_{-m}^{j}~L^{T}_{m} - L^{T}_{-m} \alpha_{m}^{j});
\label{e-cov}
\ee
we can now construct the generators of the Poincar\'e group as in (\ref{p-b}) and the physical Hilbert space 
$
{\cal H}^{(\mu,\alpha)}_{\rm phys}
$ 
as in (\ref{phys-a1}); we take 
$
D = 26
$
such that the Poincar\'e algebra closes. The Hilbert space bundle 
$
{\cal B}^{[\mu,\alpha]}
$
is made of couples
$
(p,f)
$
where 
$
p = (p^{+},\tilde{p}) \in \R_{+} \otimes \R^{D-2}
$
is a chart on the mass-shell and
$
f \in {\cal H}^{(\mu,\alpha)}_{\rm phys};
$
there is a natural fibration over the mass shell given by the canonical projection on the first component. On this bundle we have the following action of the Lorentz algebra:
\be
\xi \cdot (p,f) = (\xi\cdot p, \xi\cdot f) \quad \forall \xi \in {\rm Lie}({\cal L})
\ee
where
\be
j^{\mu\nu}\cdot p = L^{\mu\nu} \cdot p
\label{kin}
\ee
and
\be
j^{kl} = J^{(\alpha)kl}
\quad
j^{k+} = 0
\quad
j^{k-} = {1\over p^{+}} E^{k}(p)
\quad
j^{+-} = 0;
\label{p-b-bundle}
\ee
here
$
j,k = 1,\dots,D-2.
$
The scalar product in the fiber over $p$ is simply the scalar product from
$
{\cal F}^{(\alpha)}.
$
It is easy to verify all the axioms of a Hilbert space bundle. As it is well-known the representations of the Poincar\'e group are induced by representations of the stability subgroup of any point on the mass-shell. If we take the point
$
p^{(0)}
$
with coordinates
$
p^{+} = {\mu\over \sqrt{2}}, p^{j} = 0 \quad (j = 1,\cdots,D-2)
$
it is easy to get from (\ref{kin}) that the stability subgroup is 
$
SO(D-1)
$
and the infinitesimal generators are 
$
j^{kl} \quad k,l = 1,\dots,D-1.
$
Next we get from (\ref{p-b-bundle}) that the representation of 
$
SO(D-1)
$
inducing the representation of the Poincar\'e group from the theorem is
\be
j^{kl} = J^{(\alpha)kl}
\quad
j^{k+} = 0
\quad
j^{k-} = - {1\over \mu} E^{j}(p^{(0)})
\quad
j^{+-} = 0;
\ee
one can check the representation property using the definition of
$
{\cal H}^{(\mu,\alpha)}_{\rm phys}.
$

We give now the covariant description of the preceding construction. We define the Hilbert space bundle 
$
{\cal B}^{[\mu,\alpha]}_{\rm cov}
$
of couples
$
(p,\Psi)
$
where $p$ is on the positive mass-shell
$
p \in \R^{D} \quad p^{0} > 0 \quad p^{2} = \mu^{2}
$
and
$
\Psi \in {\cal F}^{[\mu,\alpha]}_{\rm cov}
$
verifies the supplementary restrictions
\be
L_{m} \Psi = 0 \quad m \geq  0
\ee
where the Virasoro operators
$
L_{m} = L_{m}^{(\alpha)}
$
are given by (\ref{vir-cov}) for the value
$
a = 1.
$
The Hermitian form in the fiber over $p$ is the form defined on
$
{\cal F}^{(\alpha)}_{\rm cov}.
$

We want to obtain an isomorphism to the previously obtained
$
{\cal H}_{phys}.
$
We present briefly the usual argument with some simplifications. We first define the 
{\it DDF states} as linear combinations of states of the form:
\be
f = A^{j_{1}}_{m_{1}} \dots A^{j_{l}}_{m_{l}}\Omega \quad m_{1},\dots,m_{l} < 0. 
\ee

It is useful to introduce the notation
\be 
K_{m} \equiv k\cdot \alpha_{m}
\ee
and we easily obtain
\be
[ K_{m}, K_{n} ] = 0 \qquad [K_{m}, L_{n} ] = m~K_{m+n};
\label{kl}
\ee
we also have for any DDF state:
\be
K_{m} f = 0 \quad \forall m > 0.
\ee
Next we have the following technical result for which we present a simpler proof:
\begin{prop}
The vectors of the type
\be
\Psi_{\lambda,\mu,f} \equiv L^{\lambda_{1}}_{-1} \cdots L^{\lambda_{m}}_{-m}
K^{\mu_{1}}_{-1} \cdots K^{\mu_{n}}_{-n}~f
\ee
where
$
\lambda_{1},\dots,\lambda_{m},\mu_{1},\dots,\mu_{n} \in \N^{*}
$
and $f$ is a DDF state, are linearly independent and generates the whole space
$
{\cal F}^{(\alpha)}_{\rm cov}.
$ 
\label{basis}
\end{prop}
{\bf Proof:} (i) We know that the Hilbert space 
$
{\cal F}^{(\alpha)}_{\rm cov}
$ 
is generated by the operators
$
\alpha^{\mu}_{-m}, \quad m > 0, \quad \mu = 0,\dots,D-1
$
applied on the vacuum. It is convenient to work with the operators
\be
\alpha^{\pm}_{m} \equiv {1\over \sqrt{2}} (\alpha^{0}_{m} \pm \alpha^{D-1}_{m})
\qquad 
\alpha^{j}_{m} \quad j = 1,\dots,D-2.
\ee
If we take
$
k \in \R^{D}
$
as in the construction of the DDF operators i.e.
$
k^{+} = 0,~k^{j} = 0~(j = 1,\dots,D-2)
$
we have
\be
K_{m} = k^{-} \alpha^{+}_{m}.
\ee

So if we apply on the vacuum operators of the form
$
K_{-1},\dots,K_{-m}
$
we obtain all the states of the form 
$
\Psi^{+} = P(\alpha^{+}_{-1},\dots,\alpha^{+}_{-m})\Omega
$
with $P$ a polynomial. Now we easily compute
\be
A^{j}_{-1}\Omega = \alpha^{j}_{-1}\Omega + p^{j} V_{-1}(k)\Omega.
\ee
Because the second vector is of the type
$
\Psi^{+}
$
we can generate the states 
$
\alpha^{j}_{-1}\Omega
$
using the DDF operators and the 
$
K_{-m}
$
operators. Afterwards, using the $K$ operators we can generate all the states of the form
$
\alpha^{j}_{-1}P(\alpha^{+}_{-1},\dots,\alpha^{+}_{-m})\Omega.
$
Using now 
$
2,3,\dots
$
DDF operators  we can establish by inductions that all states of the form 
$
P(\alpha^{j}_{-1},\alpha^{+}_{-1},\dots,\alpha^{+}_{-m})\Omega
$
can be obtained using only DDF and $K$ operators. Next we suppose that we can
create all states of the form 
$
P(\alpha^{j_{1}}_{-1},\dots,\alpha^{j_{n-1}}_{-(n-1)},
\alpha^{+}_{-1},\dots,\alpha^{+}_{-m})\Omega
$
using only DDF and $K$ operators and extend the result to $n$ by applying DDF operators
of the form 
$
A^{j}_{-n}
$
on such a state. Finally, we note that we have
\be
L^{\lambda_{1}}_{-1} \cdots L^{\lambda_{m}}_{-m} 
= {\rm const} (\alpha^{-}_{-1})^{\lambda_{1}+\cdots \lambda_{m}} + \cdots
\ee
where by $\cdots$ we mean terms containing
$
\alpha^{-}_{-1}
$
at a power strictly smaller that 
$
\lambda_{1}+\cdots \lambda_{m}.
$
If we choose the preceding sum conveniently we can generate all states with 
$
\alpha^{-}_{-1}
$
factors. In the same way we obtain the states with 
$
\alpha^{-}_{-l} \quad l > 1
$
factors. It follows that the states of the form 
$
\Psi_{\lambda,\mu,f}
$
are really generating the whole Hilbert space 
$
{\cal F}^{(\alpha)}_{\rm cov}.
$ 

(ii) We must prove that there are no linear dependencies between such vectors. If we use
the well-known relation 
\be
[ \bar{L}_{0}, \alpha^{\mu}_{-m} ] = m~\alpha^{\mu}_{-m} 
\ee 
we easily obtain that the vector 
\be
\Psi = \prod_{n,\rho} (\alpha^{\rho}_{-n})^{\epsilon_{n,\rho}}\Omega,
\label{vect1}
\ee
where the product runs over a finite set of indices, is an eigenvector of
$
\bar{L}_{0}
$
corresponding to the eigenvalue
\be
\lambda = \sum_{n,\rho} n \epsilon_{n,\rho}.
\ee
We will denote by
$
{\cal F}_{n}
$
the eigenspace of
$
\bar{L}_{0}
$
corresponding to the eigenvalue $n$. One can prove that
\be
{\cal F}_{m} \cap {\cal F}_{n} = \{ 0 \}
\ee
for 
$
m \not= n
$
using a Vandermonde determinant. Because the subspaces
$
{\cal F}_{n}, \quad n \geq 0
$
generate the whole 
$
{\cal F}^{(\alpha)}_{\rm cov}
$ 
we have the direct sum decomposition 
\be
{\cal F}^{(\alpha)}_{\rm cov} = \oplus_{n \geq 0} {\cal F}_{n}.
\ee

Now we use the relations
\be
[ \bar{L}_{0}, L_{-m} ] = m~L_{-m} \qquad
[ \bar{L}_{0}, K_{-m} ] = m~K_{-m} \qquad
[ \bar{L}_{0}, A^{j}_{-m} ] = m~A^{j}_{-m}
\ee
and find out that the vector 
\be
\Psi^{\prime} = \prod L_{-m}^{\lambda_{m}} \prod K_{-n}^{\mu_{n}}
\prod (A^{j}_{-p})^{\beta_{p,j}}\Omega
\label{vect2}
\ee
is an eigenvector of
$
\bar{L}_{0}
$
corresponding to the eigenvalue
\be
\lambda^{\prime} = \sum_{n} n (\lambda_{n} + \mu_{n} + \sum_{j} \beta_{n,j}).
\ee

Let us fix
$
N \in \N.
$
Then
$
{\cal F}_{N}
$
is generated by vectors of the type (\ref{vect1}) with 
$
\lambda = N;
$
on the other hand the vectors of the type (\ref{vect2}) are also generating the whole Hilbert space but only those corresponding to
$
\lambda^{\prime} = N
$
are in
$
{\cal F}_{N}.
$
It follows that
$
{\cal F}_{N}
$
is generated by the vectors of the type (\ref{vect2}) corresponding to
 $
\lambda^{\prime} = N.
$
Because the (finite) number of vectors of the type (\ref{vect1}) corresponding to 
$
\lambda = N
$
is the same as the number of the vectors of the type (\ref{vect2}) corresponding to 
$
\lambda^{\prime} = N
$
it means that the vectors of the type (\ref{vect2}) corresponding to 
$
\lambda^{\prime} = N
$
must be linear independent.
$\qed$

We note that in our proof we did not have to compute the complicated determinant used in the proof from \cite{GSW}.

The rest of the proof is standard and can be found in \cite{GSW}. The final result is:
\begin{prop}
Let
$
D = 26
$ 
and the vector 
$
\Psi \in {\cal F}^{(\alpha)}_{\rm cov}
$
verifying
\be
L_{m}\Psi = 0 \quad \forall m \geq 0;
\ee
then we can uniquely write it in the form 
\be
\Psi = f + s
\ee 
where $f$ is a DDF state,  
$
s \in S
$ 
and we have
\be
L_{0} f = f \qquad L_{0} s = s \qquad L_{m} s = 0 \quad (\forall m > 0).
\label{fs}
\ee
\end{prop}
 
The end of this analysis is:
\begin{thm}
The Hermitian form on the Hilbert space bundle
$
{\cal B}^{[\mu,\alpha]}_{\rm cov}
$
is positively defined. If we factor out the states of null norm we obtain a representation of the Poincar\'e group equivalent to the representation in the bundle
$
{\cal B}^{[\mu,\alpha]}.
$
\end{thm}
{\bf Proof:} (i) If we use the preceding proposition we can write any element in the fiber over $p$ as
$
\Psi = f + s
$ 
where $f$ is a DDF state,
$
s \in S
$ 
and we have the relations (\ref{fs}). From these relations it easily follows that we have
\be
< \Psi,\Psi>~=~<f,f>~~\geq 0
\ee
so if we eliminate the null-norm states we end up with a factor Hilbert space bundle with fibres isomorphic to the subspace of DDF states.

(ii) We have to determine the representation of the stability subgroup
$
SO(D-1)
$
of the point
$
p^{(0)}.
$
It is clear that we have
\be
J^{(\alpha)kl}~A^{j}_{n} = i~(\delta_{kj}~A^{l}_{n} - \delta_{lj}~A^{k}_{n}) \quad
j,k,l = 1,\dots,D-2
\ee
so we have for any DDF state 
\be
J^{(\alpha)kl}f = j^{(\alpha)kl}f.
\label{ddf0}
\ee

We still have to compute the action of the generators
$
J^{(\alpha)k,D-1}
$
on the fiber. It is important to note that from the first relation (\ref{fs}) we have
\be
\alpha^{-}_{m}f = {\sqrt{2}\over \mu}~\bar{L}_{m}f \quad \forall m > 0;
\ee
also it is easy to prove that
\be
\alpha^{+}_{m}f = 0 \quad \forall m > 0.
\ee

Using these relations it follows that for any two DDF states 
$
f, f^{\prime}
$
we have
\be
<f^{\prime}, J^{(\alpha)k,D-1}f> = <f^{\prime}, j^{(\alpha)k,D-1}f>
\ee
where in the right-hand side we have the operators (\ref{p-b-bundle}). It is more complicated to extend this relation for 
$
f^{\prime} \rightarrow \Psi \in {\cal F}^{(\alpha)}_{\rm cov};
$
for this we have to use the generic form of states 
$
\Psi_{\lambda,\mu,f} 
$
and commute 
$
L_{m}
$
and
$
K_{m}
$
with 
$
E^{j}(p^{(0)}).
$

As a result we have for any DDF state
\be
J^{(\alpha)k,D-1}f = j^{(\alpha)k,D-1}f.
\label{ddf1}
\ee

Next we note that we have for any DDF state
\be
\bar{L}_{0} f = \left(1 + {\mu^{2}\over 2} \right)f;
\ee
from here it follows that
\be
<f^{\prime},\bar{L}^{T}_{0} f> = \left(1 + {\mu^{2}\over 2} \right)<f^{\prime},f>
\ee
where
$
\bar{L}^{T}_{0} = L^{T}_{0}
$
is the transversal part of
$
\bar{L}_{0}
$
(i.e. it contains only the modes
$
1,\dots,D-2
$).

As above we can extend the relation for
$
f^{\prime} \rightarrow \Psi \in {\cal F}^{(\alpha)}_{\rm cov}
$
so we have
\be
L^{T}_{0} f = \left(1 + {\mu^{2}\over 2} \right)f
\label{ddf2}
\ee
for any DDF state. 
From (\ref{ddf1}) and (\ref{ddf2}) it follows that the fiber over
$
p^{(0)}
$
of the fiber bundle
$
{\cal B}^{[\mu,\alpha]}_{\rm cov}
$
coincides with the fiber over the same point of the fiber bundle
$
{\cal B}^{[\mu,\alpha]}.
$
According to a standard theorem 9.20 of \cite{Va} it follows that the two representations of the Poincar\'e group are equivalent.
$\qed$
\section{BRST Quantization of the Bosonic String\label{brst}}

Another possibility is to introduce ghost degrees of freedom and construct a gauge charge operator $Q$ which squares to zero
$
Q^{2} = 0
$
in such a way that there is a canonical isomorphism between the physical
Hilbert space and the factor space
$
Ker(Q)/Im(Q)
$
\cite{KO}, \cite{FO}, \cite{T}, \cite{BP}, \cite{FGZ} \cite{P}. We provide here an elementary treatment. First we define the ghost Hilbert space
$
{\cal F}^{gh}_{1};
$
by definition it is generated by the operators
$
b_{m}, c_{m} \quad m \in \Z
$ 
from the vacuum
$\Omega_{gh} \in {\cal F}^{gh}_{1}$;
we assume that
\be
b_{m}\Omega_{gh} = 0 \quad c_{m}\Omega_{gh} = 0 \quad \forall m > 0.
\ee

These operators are subject to the following anticommutation relations:
\be
\{b_{m}, b_{n}\} = 0 \quad \{c_{m}, c_{n}\} = 0 
\quad \{b_{m}, c_{n}\} = \delta_{m+n} \cdot I;
\ee 
we also suppose that there is a conjugation operation in
$
{\cal F}^{gh}_{1}
$
such that
\be
b_{m}^{\dagger} = b_{-m} \quad c_{m}^{\dagger} = c_{-m}.
\ee

We can give a concrete realization as follows:
$
{\cal F}^{gh}_{1} = {\cal F}_{b,c} \otimes {\cal C}
$
where 
$
{\cal F}_{b,c}
$
is the Fock space generated by the operators
$
b_{m}, c_{m} \quad m \in \Z^{*}
$ 
and
$
{\cal C}
$
is the Clifford algebra generated by 
$
b_{0}, c_{0}.
$

\begin{prop}
The following operators 
\be
l^{(1)}_{m} = \sum_{n \in \Z} (m+n) :b_{m-n}c_{n}:
\ee
are well defined on the algebraic Hilbert space and are verifying:
\bea
[ l^{(1)}_{m}, b_{n}] = (m-n) b_{m+n} \qquad
[ l^{(1)}_{m}, c_{n}] = - (2m+n) b_{m+n}
\nonumber \\
~[ l^{(1)}_{m}, l^{(1)}_{n}] = (m-n) l^{(1)}_{m+n} 
+ {1\over 6} m(1 - 13m^{2}) \delta_{m+n} \cdot I
\nonumber \\
(l^{(1)}_{m})^{\dagger} = l^{(1)}_{-m}.
\eea
\end{prop}
{\bf Proof:} We write
\be
l^{(1)}_{m} = l_{m}^{\prime} + m b_{m} c_{0} + 2 m b_{0} c_{m}
\ee
where 
$
l_{m}^{\prime}
$
contains only the non-zero modes:
\be
l_{m}^{\prime} = \sum_{n \not= 0,m} (m+n) :b_{m-n}c_{n}:
\ee 

For the non-zero modes the $2$-point functions are
\bea
<\Omega_{gh},b_{m}c_{n}\Omega_{gh}> = \theta(m) \delta_{m+n} \quad
<\Omega_{gh},c_{m}b_{n}\Omega_{gh}> = \theta(m) \delta_{m+n} \quad
\nonumber \\
<\Omega_{gh},b_{m}b_{n}\Omega_{gh}> = 0 \quad
<\Omega_{gh},c_{m}c_{n}\Omega_{gh}> = 0
\eea
and we can compute the commutators from the statement using Wick theorem.
$\qed$

Next we have
\begin{cor}
Let us consider in the Hilbert spaces
$
{\cal H} \equiv {\cal H}^{[\mu,\alpha]}_{cov} \otimes {\cal F}^{gh}_{1}
$
the following operators:
$
L_{m}^{(\alpha)} 
$ 
cf. (\ref{vir-cov}) and
\be
{\cal L}_{m}^{(\alpha)} = L_{m}^{(\alpha)} \otimes I_{2} + I_{1} \otimes l_{m}^{(1)};
\ee
then we have
\bea
[ {\cal L}_{m}^{(\alpha)}, {\cal L}_{n}^{(\alpha)}] = (m-n) {\cal L}_{m+n}^{(\alpha)}
+ m \left( {D-26\over 12} m^{2} + 2a - {D-2\over 12} \right) \delta_{m+n} \cdot I
\nonumber \\
({\cal L}_{m}^{(\alpha)})^{\dagger} = {\cal L}_{-m}^{(\alpha)}.
\eea 
\end{cor}
In this enlarged Hilbert space we have \cite{KO}:
\begin{prop}
The following operator
\be
Q \equiv \sum L^{(\alpha)}_{-m} c_{m} - {1\over 2} \sum (m-n) :c_{-m} c_{-n} b_{m+n}:
\ee
is well defined on the algebraic Hilbert space and is formally self-adjoint; it verifies
\be
Q^{2} = 0
\ee
{\it iff}
$
D = 26
$
and
$a = 1$.
\end{prop}
{\bf Proof:} We separated the non-zero modes as before:
\be
Q = Q_{0} + {\cal L}_{0}^{(\alpha)} c_{0} + C^{(1)}_{0} b_{0}
\ee
where
\be
Q_{0} \equiv \sum_{m \not= 0} {\cal L}^{(\alpha)}_{-m} c_{m} 
- {1\over 2} \sum_{m,n \not= 0} \sum_{m+n \not= 0} (m-n) :c_{-m} c_{-n} b_{m+n}:
\ee
and
\be
C^{(1)}_{m} \equiv {1\over 2} \sum_{p+q=m} (p - q) :c_{p} c_{q}:
= {1\over 2} \sum_{p+q=m} (p - q) c_{p} c_{q}
\ee

The most convenient way to prove the theorem is the following. One proves by direct computation (using our preferred method based on Wick theorem) the following formulas:
\bea
\{Q, b_{m} \} = {\cal L}_{m}^{(\alpha)} \qquad
\{ Q, c_{m} \} = C^{(1)}_{m} 
\nonumber \\
~[ Q, {\cal L}_{m}^{(\alpha)} ] = \rho_{m} c_{m} \qquad
~[ Q, K_{m} ] = - m \sum K_{m-n} c_{n}
\label{Q-b1}
\eea
where
\be
\rho_{m} \equiv - m \left( {D-26\over 12} m^{2} + 2a - {D-2\over 12}\right).
\ee

We now use the following observation. According to proposition \ref{basis} we can take in
${\cal H}$
the following basis:
\be
\Psi^{\prime} = \prod b_{-i} \prod c_{-j} \prod L^{(\alpha)}_{-m} \prod K_{-n}~f
\ee
where $f$ are DDF states, the indices of type $m, n$ are strictly positive and the indices of the type $i,j$ are 
$\geq 0$.
It is easy to substitute
$
L^{(\alpha)}_{m} = {\cal L}_{m}^{(\alpha)} - l_{m}^{(1)}
$
and consider the new basis
\be
\Psi = \prod b_{-i} \prod c_{-j} \prod {\cal L}^{(\alpha)}_{-m} \prod K_{-n}~f
\label{basis-gh}
\ee

Because
$
{\cal L}_{m}^{(\alpha)}~f = 0\quad \forall m \geq 0
$
we easily find out that
\be
Qf = 0
\label{Q-b2}
\ee
for any DDF state $f$. The operator $Q$ is perfectly well defined by (\ref{Q-b1}) and 
(\ref{Q-b2}); indeed we know how to act with $Q$ on states of the form (\ref{basis-gh}): we
commute $Q$ using (\ref{Q-b1}) till it hits a DDF state and gives $0$ according to 
(\ref{Q-b2}). Now it is easy to obtain from (\ref{Q-b1}):
\be
\{Q^{2}, b_{m} \} = \rho_{m} b_{m} \qquad
\{ Q^{2}, c_{m} \} = 0 \qquad
[ Q^{2}, {\cal L}_{m}^{(\alpha)} ] = \rho_{m} C^{(1)}_{m} \qquad
[ Q^{2}, K_{m} ] = 0.
\label{Q-b3}
\ee
Because we obviously have
$
Q^{2}f = 0
$
it immediately follows that
\be
Q^{2} = 0 \Longleftrightarrow \rho_{m} = 0 \Longleftrightarrow D = 26, a = 1 
\label{Q-b5}
\ee 
i.e. the statement of the theorem.
$\qed$
\begin{rem}
One can directly prove that 
\be
Q^{2} = {1\over 2} \sum m \left( {D-26\over 12} m^{2} + 2a - {D-2\over 12}\right)
:c_{-m} c_{m}:
\ee
which is another way to obtain the result. Let us also note that if the conditions
$
D = 26, a = 1 
$
are meet then we also have no anomalies in the Virasoro algebra:
\be
[ {\cal L}_{m}^{(\alpha)}, {\cal L}_{n}^{(\alpha)}] = (m-n) {\cal L}_{m+n}^{(\alpha)}.
\ee
\end{rem}

To analyze the cohomology of the BRST operator $Q$ we need the following result:
\begin{prop}
The operator 
$\tilde{Q}$
is well defined on the algebraic Hilbert space through the following formulas:
\be
\{\tilde{Q}, b_{m} \} = 0 \qquad
\{ \tilde{Q}, c_{m} \} = \delta_{m} \cdot I \qquad
[ \tilde{Q}, {\cal L}_{m}^{(\alpha)} ] = - m b_{m} \qquad
[ \tilde{Q}, K_{m} ] = 0
\label{tQ-b1}
\ee
and
\be
\tilde{Q}f = 0
\ee
for any DDF state $f$. We also have
\be
\tilde{Q}^{\dagger} = \tilde{Q} \qquad \tilde{Q}^{2} = 0.
\ee
\end{prop}

{\bf Proof:} Because the operators 
$
b_{m}, c_{m}, {\cal L}_{m}^{(\alpha)}, K_{m} 
$
are connected by various relations, we have to verify the Jacobi identities of the type:
\be
[[X,Y], \tilde{Q} ]_{\rm graded} + {\rm cyclic~permutations} = 0
\ee
where
$X, Y$
are operators from the set
$
b_{m}, c_{m}, {\cal L}_{m}^{(\alpha)}, K_{m}.
$ 
The non-trivial ones are corresponding to the pairs
$
({\cal L}_{m}^{(\alpha)}, {\cal L}_{n}^{(\alpha)})
$
and
$
({\cal L}_{m}^{(\alpha)}, c_{n})
$
and are easily checked.
$\qed$

The main result is the following
\begin{thm}
If 
$
\Psi \in {\cal H}
$
verifies
$
Q \Psi = 0
$
then it is of the form 
\be
\Psi = Q \Phi + f_{1} + b_{0} f_{2} + c_{0} f_{3}
\ee
where 
$
f_{j}
$
are DDF states. 
\end{thm}
{\bf Proof:} 
A good strategy to determine the cohomology of the operator $Q$ is to mimic Hodge theorem i.e. to find a homotopy operator 
$
\tilde{Q}
$
such that the spectrum of the ``Laplacian"
\be
\Delta \equiv Q\tilde{Q} + \tilde{Q}Q
\ee
can be easily determined. We take such an operator to be the one determined in the previous proposition. It is now elementary to see that the Laplace operator is alternatively given by:
\bea
[ \Delta, b_{m} ] = - m b_{m} \quad [ \Delta, c_{m} ] = - m c_{m} \qquad
[ \Delta, {\cal L}^{(\alpha)}_{m} ] = - m {\cal L}^{(\alpha)}_{m} \quad 
[ \Delta, K_{m} ] = - m K_{m} 
\nonumber \\
\Delta f = 0.
\eea

It follows that we have on states of the form (\ref{basis-gh})
\be
\Delta \Psi = (\sum m + \sum n + \sum i + \sum j) \Psi;
\label{delta}
\ee 
we observe that the eigenvalues from (\ref{delta}) are 
$\geq 0$
the equality sign being true only for vectors of the form 
$
f_{1} + b_{0} f_{2} + c_{0} f_{3} 
$
with
$
f_{j}
$
DDF states. It means that every vector in
${\cal H}$
is of the form 
\be
\Psi = \Psi_{0} + f_{1} +b_{0} f_{2} + c_{0} f_{3}
\ee
where 
$
\Psi_{0}
$
belongs to the subspace
$
{\cal H}^{\prime} \subset {\cal H}
$
of vectors with strictly positive eigenvalues of 
$
\Delta.
$

Now suppose that the vector 
$\Psi$
verifies the equation from the statement
$
Q\Psi = 0.
$
Then it is easy to see that we also have
$
Q\Psi_{0} = 0
$
so
$
\Delta\Psi_{0} = Q \tilde{Q} \Psi_{0}.
$
We can write 
$
\Psi_{0} = \sum \Psi_{\omega}
$
where 
$
\Psi_{\omega}
$
are linear independent vectors from
${\cal H}^{\prime}$
corresponding to distinct eigenvalues
$\omega > 0$.
Then the preceding relation is equivalent to
$
\Psi_{\omega} = {1\over\omega}~Q \tilde{Q} \Psi_{\omega};
$
if we define
$
\Phi = \sum {1\over \omega}~\tilde{Q} \Psi_{\omega}
$
then it follows that
$
\Psi_{0} = Q \Phi
$
and this finishes the proof.
$\qed$

We conclude that the cohomology of the operator $Q$ is described by three copies of the physical space of DDF states. To eliminate this tripling we proceed as follows. We construct the Hilbert space 
${\cal F}^{gh}_{1}$
such that it also verifies 
\be
b_{0} \Omega_{gh} = 0;
\ee
in this case 
${\cal F}^{gh}_{1}$
is a Fock space. Then we construct ${\cal H}$ as above and consider the subspace
\be
{\cal H}_{0} \equiv \{ \Psi \in {\cal H} | b_{0}\Psi = 0 \quad 
{\cal L}^{(\alpha)}_{0} \Psi = 0 \}.
\ee

This subspace is left invariant by the operator $Q$; if
$
\Psi \in {\cal H}_{0}
$
verifies
$
Q\Psi = 0
$
then we have similarly with the preceding theorem
\be
\Psi = Q\Phi + f 
\ee 
with $f$ some DDF state. 

\section{The Covariant Neveau-Schwarz Superstring}

We proceed as before from the Hilbert space
$
{\cal H}^{(NS)}_{\rm cov}
$
generated by the operators 
$
\alpha^{\mu}_{m}, m \in \Z, \mu = 0,\dots,D
$
and
$
b^{\mu}_{r}, r \in {1\over 2} + \Z, \mu = 0,\dots,D
$
such that:
\bea
[ \alpha^{\mu}_{m}, \alpha^{\nu}_{n} ] = - \eta_{\mu\nu}~m~\delta_{m+n} \cdot I, 
\quad \forall m, n
\nonumber \\
\alpha^{\mu}_{m} \Omega = 0, \quad m > 0
\nonumber \\
(\alpha^{\mu}_{m})^{+} = \alpha^{\mu}_{-m} \quad \forall m;
\nonumber \\
\{ b^{\mu}_{r}, b^{\nu}_{s} \} = - \eta_{\mu\nu}~m~\delta_{r+s} \cdot I, 
\quad \forall r, s
\nonumber \\
b^{\mu}_{r} \Omega = 0, \quad r > 0
\nonumber \\
(b^{\mu}_{r})^{+} = b^{\mu}_{-r} \quad \forall r
\eea
and define the covariant Virasoro operators
\be
L^{(NS)}_{m} \equiv - {1\over 2}~\eta_{\mu\nu}~\sum_{n \in \Z}~
:\alpha^{\mu}_{m-n} \alpha^{\nu}_{n}:
- {1\over 2}~\eta_{\mu\nu}~\sum_{r \in 1/2+\Z}~r~:b^{\mu}_{-r} b^{\nu}_{m+r}: 
- a~\delta_{m} \cdot I
\label{vir-NS}
\ee
and the supersymmetric partners
\be
G_{r} \equiv - \eta_{\mu\nu}~\sum_{n \in \Z} \alpha^{\mu}_{-n} b^{\nu}_{n+r}
\ee
such that we have the following relations:
\bea
[ L^{(NS)}_{m}, L^{(NS)}_{n} ] = (m - n) L^{(NS)}_{m+n} 
+ \left[ D~{ m (m^{2} - 1) \over 8} + 2 m a \right]~\delta_{m+n}~\cdot I. 
\nonumber \\
~[ L^{(NS)}_{m}, G_{r} ] = \left({m\over 2} - r\right) G_{m+r}
\nonumber \\
\{ G_{r}, G_{s} \} = 2 L^{(NS)}_{r+s} 
+ \left[ {D\over 2}~\left(r^{2} - {1\over 4}\right) + 2 a \right]~\delta_{r+s}~\cdot I
\nonumber \\
(L^{(NS)}_{m})^{\dagger} = L^{(NS)}_{-m} \qquad G_{r}^{\dagger} = G_{-r}.
\eea
In this Hilbert space we have an action of the supersymmetric Poincar\'e algebra. Then we can obtain the Neveau-Schwarz case if we take 
$D = 10,~a = 1/2$
and restrict the states by the conditions:
\be
L_{m}^{(NS)} \Psi = 0 \quad \forall m \geq 0 \qquad
G_{r} \Psi = 0 \quad \forall r > 0.
\ee

The DDF states can be constructed as before. First we have to construct operators
$
A^{j}_{m},~B^{j}_{r}, j = 1,\dots,D-1
$
such that they verify the same algebra as the operators
$
\alpha^{j}_{m}, b^{j}_{r}
$
and they commute with
$
G_{r},\forall r;
$
(this implies that they commute with 
$
L_{m}, \forall m.)
$
The DDF states are generated by these operators from the vacuum. 

For the BRST description we need to enlarge the ghost space: we consider the Fock space
$
{\cal F}^{gh}_{2}
$
generated by the operators
$
\beta_{r}, \gamma_{r} \quad r \in {1\over 2} + \Z
$ 
from the vacuum
$\Omega_{gh} \in {\cal F}^{gh}_{2}$;
we assume that
\be
\beta_{r}\Omega_{gh} = 0 \quad \gamma_{r}\Omega_{gh} = 0 \quad \forall r > 0.
\ee

These operators are subject to the following anticommutation relations:
\be
[\beta_{r}, \beta_{s} ] = 0 \quad [ \gamma_{r}, \gamma_{s} ] = 0 \quad 
[ \gamma_{r}, b_{s} ] = \delta_{r+s} \cdot I;
\ee 
we also suppose that there is a conjugation operation in
$
{\cal F}^{gh}_{2}
$
such that
\be
\beta_{r}^{\dagger} = \beta_{-r} \quad \gamma_{r}^{\dagger} = \gamma_{-r}.
\ee

We can define as usual the algebraic Hilbert space (the subspace of vectors generated by a finite number of operators
$
\beta_{r}, \gamma_{r} \quad r \leq 0)
$ 
and normal ordering in
$
{\cal F}^{gh}_{2}.
$
\begin{prop}
The following operators 
\be
l^{(2)}_{m} = \sum_{r \in 1/2 + \Z} \left({m\over 2}+r\right) :\beta_{m-r}\gamma_{r}:
\ee
are well defined on the algebraic Hilbert space and are verifying:
\bea
[ l^{(2)}_{m}, \beta_{r}] = \left({m\over 2}-r\right) \beta_{m+r} \qquad
[ l^{(2)}_{m}, \gamma_{r}] = - \left({3m\over 2}+r\right) \gamma_{m+r}
\nonumber \\
~[ l^{(2)}_{m}, l^{(2)}_{n}] = (m-n) l^{(2)}_{m+n} 
+ {1\over 12} m(11m^{2} + 1) \delta_{m+n} \cdot I
\nonumber \\
(l^{(2)}_{m})^{\dagger} = l^{(2)}_{-m}.
\eea
\end{prop}
{\bf Proof:} The $2$-point functions are
\bea
<\Omega_{gh},\beta_{r}\gamma_{s}\Omega_{gh}> = -\theta(r) \delta_{r+s} \quad
<\Omega_{gh},\gamma_{r}\beta_{s}\Omega_{gh}> = \theta(r) \delta_{r+s} \quad
\nonumber \\
<\Omega_{gh},\beta_{r}\beta_{s}\Omega_{gh}> = 0 \quad
<\Omega_{gh},\gamma_{r}\gamma_{s}\Omega_{gh}> = 0
\eea
and we can compute the commutators from the statement using Wick theorem.
$\qed$

Next we have
\begin{cor}
Let us consider in the Hilbert spaces
$
{\cal F}^{gh}_{NS} \equiv {\cal F}^{gh}_{1} \otimes {\cal F}^{gh}_{2}
$
the following operators 
\be
l^{(NS)}_{m} = l^{(1)}_{m} \otimes I_{2} + I_{1} \otimes l^{(2)}_{m};
\ee
then we have
\bea
[ l_{m}^{(NS)}, l_{n}^{(NS)}] = (m-n) l_{m+n}^{(NS)}
+ {1\over 4}~m (1 - 5m^{2}) \delta_{m+n} \cdot I
\nonumber \\
(l_{m}^{(NS)})^{\dagger} = l_{-m}^{(NS)}.
\eea 
\end{cor}

Next, we consider in
$
{\cal F}^{gh}_{NS}
$
the following operators
\be
g_{r} \equiv - 2 \sum_{n \in \Z} b_{-n} \gamma_{n+r}
+ \sum_{n \in \Z} \left( {n\over 2} - r\right) c_{-n} \beta_{n+r}
\ee
which are well-defined on the algebraic Fock space. We have
\begin{prop}
The following relations are verified 
\bea
g_{r}^{\dagger} = g_{-r}
\nonumber \\
~[l_{m}^{(NS)}, g_{r} ] = \left( {m\over 2} - r\right) g_{m+r}
\nonumber \\
\{g_{r},g_{s}\} = 2 l_{r+s}^{(NS)} + \left( {1\over 4}- 5r^{2} \right) \delta_{r+s} \cdot I.
\eea
\end{prop}

The proofs of the first two relations are elementary. For the last one we use Wick theorem.
Next we have
\begin{cor}
Let us consider in the Hilbert spaces
$
{\cal H} \equiv {\cal H}^{(b)}_{cov} \otimes^{s} {\cal F}^{gh}_{NS}
$
where the skew tensor product
$
\otimes^{s}
$
is such that we have normal (anti)commutation relations i.e. the Fermionic operators
$
b^{\mu}_{r} 
$
are anticommuting with
$
b_{m}, c_{m}.
$
We then define the operators 
\bea
{\cal L}_{m}^{(NS)} = L^{(NS)}_{m} \otimes I_{2} + I_{1} \otimes l^{(NS)}_{m}
\nonumber \\
{\cal G}_{r} = G_{r} \otimes I_{2} + I_{1} \otimes g_{r}
\eea
and we have
\bea
[ {\cal L}_{m}^{(NS)}, {\cal L}_{n}^{(NS)}] = (m-n) {\cal L}_{m+n}^{(NS)}
+ m \left( {D-10\over 8} m^{2} + 2a - {D-2\over 8} \right) \delta_{m+n} \cdot I
\nonumber \\
\{ {\cal G}_{r}, {\cal G}_{s} \} = 2 {\cal L}^{(NS)}_{r+s} 
+ \left( {D-10\over 8}~r^{2} + 2 a - {D-2\over 8}\right)~\delta_{r+s}~\cdot I
\nonumber \\
({\cal L}^{(NS)}_{m})^{\dagger} = {\cal L}^{(NS)}_{-m} \qquad 
{\cal G}_{r}^{\dagger} = {\cal G}_{-r}.
\eea 
The anomalies cancel iff
$D = 10, a = 1/2$.
\end{cor}
In this enlarged Hilbert space we have:
\begin{prop}
The following operator
\bea
Q = Q^{(NS)} \equiv \sum L^{(NS)}_{-m} c_{m} 
- {1\over 2} \sum (m-n) :c_{-m} c_{-n} b_{m+n}:
\nonumber \\
+ \sum G_{-r} \gamma_{r} + \sum c_{-m} l^{(2)}_{m}  + \sum b_{-m} C^{(2)}_{m}  
\eea
where
\be
C^{(2)}_{m} \equiv - \sum_{r+s=m} :\gamma_{r}\gamma_{s}: = 
- \sum_{r+s=m} \gamma_{r}\gamma_{s}
\ee
is well defined on the algebraic Hilbert space and it is formally self-adjoint; it verifies
\be
Q^{2} = 0
\ee
{\it iff}
$
D = 10
$
and
$a = 1/2$.
\end{prop}
{\bf Proof:} It is convenient to denote by
$
Q_{j}, j = 1,\dots,5
$
the five terms in the expression of the BRST charge and write
\be
Q = Q^{\prime} + \sum_{j=3}^{5} Q_{j}
\ee
where the sum of the first two terms
$ 
Q^{\prime}
$
can be obtained from $Q$ of the preceding Section with the substitution
$
L^{(\alpha)}_{m} \longrightarrow L^{(NS)}_{m}
$
so we can use some of the computations performed there. We introduce the notation
\be
H_{r} \equiv k_{\mu}~b^{\mu}_{r}. 
\ee
and we have as before:
\bea
\{Q, b_{m} \} = {\cal L}_{m}^{(NS)} \qquad
\{ Q, c_{m} \} = C^{(NS)}_{m} \equiv C^{(1)}_{m} + C^{(2)}_{m}
\nonumber \\
~[ Q, \beta_{r} ] = {\cal G}_{r} \qquad
[ Q, \gamma_{r} \} = - \sum \left( {3m\over 2} + r\right) c_{-m} \gamma_{m+r}
\nonumber \\
~[ Q, {\cal L}_{m}^{(NS)} ] = \rho_{m} c_{m} \qquad
\{ Q, {\cal G}_{r} \} = \lambda_{r} \gamma_{r}
\nonumber \\
~[ Q, K_{m} ] = - m \sum K_{m-n} c_{n} + \sum H_{m-r} \gamma_{r} 
\nonumber \\
\{ Q, H_{r} \} = - \sum \left( {m\over 2} - r\right) H_{r-m} c_{m} + \sum K_{r-s} \gamma_{s}.
\label{Q-ns1}
\eea
where
\be
\rho_{m} \equiv - m \left( {D-10\over 8} m^{2} + 2a - {D-2\over 8}\right)
\qquad 
\lambda_{r} \equiv {D-10\over 2} r^{2} + 2a - {D-2\over 8}.
\ee

The only really complicated computation is for the anticommutator 
$
\{ Q, {\cal G}_{r} \}.
$
One can prove that we can take in
${\cal H}$
the following basis:
\be
\Psi = \prod b_{-i} \prod c_{-j} \prod \beta_{-r} \prod \gamma_{-s}
\prod {\cal L}_{-m}^{(NS)} \prod {\cal G}_{-t} \prod K_{-n} \prod H_{-u}~f
\ee
where $f$ are DDF states, the indices of the type $m,n \in \Z$,$\quad r,s,t,u \in 1/2 + \Z$
are taking positive values and the indices of the type $i,j \in \Z$ are $\geq 0$.

Because
$
L_{m}^{(NS)} f = 0 \quad \forall m \geq 0 \qquad
G_{r} f = 0 \quad \forall r > 0
$
we easily find out that
\be
Qf = 0
\label{Q-ns2}
\ee
for any DDF state $f$. We argue as before that the operator $Q$ is well defined by (\ref{Q-ns1}) and (\ref{Q-ns2}). Now it is easy to obtain from (\ref{Q-ns1}):
\bea
[ Q^{2}, b_{m} ] = \rho_{m} c_{m} \qquad
[ Q^{2}, c_{m} ] = 0 \qquad
[ Q^{2}, \beta_{r} ] = \lambda_{r} \gamma_{r}\qquad
[ Q^{2}, \gamma_{r} \} = 0
\nonumber \\
~[ Q^{2}, {\cal L}_{m}^{(NS)} ] = \rho_{m} C^{(NS)}_{m} \qquad
[ Q^{2}, {\cal G}_{r} ] = \lambda_{r} C^{(3)}_{r} \qquad
[ Q^{2}, K_{m} ] = 0\qquad
[ Q^{2}, H_{r} ] = 0.
\label{Q-ns3}
\eea
Because we obviously have
$
Q^{2}f = 0
$
it immediately follows that
\be
Q^{2} = 0 \Longleftrightarrow \rho_{m} = 0 \quad \lambda_{r} = 0 \Longleftrightarrow 
D = 10, a = 1/2 
\label{Q-ns5}
\ee 
i.e. the statement of the theorem.
$\qed$

To analyze the cohomology of the BRST operator 
$Q$ 
we construct as before, its homotopy:
\begin{prop}
The operator 
$\tilde{Q}$
is well defined on the algebraic Hilbert space through the following formulas:
\bea
\{\tilde{Q}, b_{m} \} = 0 \qquad
\{ \tilde{Q}, c_{m} \} = \delta_{m} \cdot I \qquad
\{\tilde{Q}, \beta_{r} \} = 0 \qquad
\{ \tilde{Q}, \gamma_{r} \} = 0
\nonumber \\
~[ \tilde{Q}, {\cal L}_{m}^{(NS)} ] = - m b_{m} \qquad
\{ \tilde{Q}, {\cal G}_{r} \} = - r \beta_{r} \qquad
[ \tilde{Q}, K_{m} ] = 0 \qquad
[ \tilde{Q}, H_{r} ] = 0
\label{tQ-ns1}
\eea
and
\be
\tilde{Q}f = 0
\ee
for any DDF state $f$. We also have
\be
\tilde{Q}^{\dagger} = \tilde{Q} \qquad \tilde{Q}^{2} = 0.
\ee
\end{prop}

{\bf Proof:} We have to verify the Jacobi identities of the type:
\be
[[X,Y], \tilde{Q} ]_{\rm graded} + {\rm cyclic~permutations} = 0
\ee
where
$X, Y$
are operators from the set
$
b_{m}, c_{m}, \beta_{r}, \gamma_{r}, {\cal L}_{m}^{(NS)}, {\cal G}_{r}, K_{m}, H_{r}.
$ 
We have some non-trivial ones corresponding to pairs
$
({\cal L}_{m}^{(NS)}, {\cal L}_{n}^{(NS)}),~
({\cal L}_{m}^{(NS)}, c_{n}),~
({\cal L}_{m}^{(NS)}, {\cal G}_{r}),
({\cal G}_{r}, {\cal G}_{s})
$
and
$
({\cal G}_{r}, \gamma_{s}).
$
$\qed$

The main result is similar to the one in the previous Section: 
\begin{thm}
If 
$
\Psi \in {\cal H}
$
verifies
$
Q\Psi = 0
$
then it is of the form 
\be
\Psi = Q\Phi + f_{1} + b_{0} f_{2} + c_{0} f_{3}
\ee
where 
$
f_{j}
$
are DDF states. 
\end{thm}
{\bf Proof:} 
The ``Laplacian" is as before
$
\Delta \equiv Q\tilde{Q} + \tilde{Q}Q
$
It is now elementary to determine the alternative expression:
\bea
[ \Delta, b_{m} ] = - m b_{m} \quad [ \Delta^{(NS)}, c_{m} ] = - m c_{m} \quad
[ \Delta, \beta_{r} ] = - r \beta_{r} \quad 
[ \Delta, \gamma_{r} ] = - r \gamma_{r} \quad
\nonumber \\
~[ \Delta, {\cal L}^{(NS)}_{m} ] = - m {\cal L}^{(NS)}_{m} \quad 
[ \Delta, {\cal G}_{r} ] = - r {\cal G}_{r} \quad
[ \Delta, K_{m} ] = - m K_{m} \quad
[ \Delta, H_{r} ] = - r H_{r}
\nonumber \\
\Delta f = 0.
\eea

It follows that we have
\be
\Delta \Psi = (\sum m + \sum n + \sum i + \sum j + \sum r + \sum s + \sum t + \sum u ) \Psi;
\label{delta-ns}
\ee 
we observe that the eigenvalues from (\ref{delta}) are 
$\geq 0$
the equality sign being true only for vectors of the form 
$
f_{1} + b_{0} f_{2} + c_{0} f_{3} 
$
with
$
f_{j}
$
DDF states, as in the previous Section. From now on the argument is the same as there.
$\qed$

To eliminate this tripling we proceed, again, as in the previous Section: we construct the Fock space 
${\cal F}^{gh}_{NS}$
such that it also verifies 
\be
b_{0} \Omega_{gh} = 0.
\ee
Then we construct ${\cal H}$ as above and consider the subspace
\be
{\cal H}_{0} \equiv \{ \Psi \in {\cal H} | \quad b_{0}\Psi = 0 \quad 
{\cal L}^{(NS)}_{0} \Psi = 0 \}.
\ee

This subspace is left invariant by the operator $Q$ and if
$Q^{(NS)}\Psi = 0$
then we have similarly with the preceding theorem
\be
\Psi = Q^{(NS)}\Phi + f 
\ee 
with $f$ some DDF state. 

\section{The Quantum Ramond Superstring}

The modification with respect to the preceding Section are minimal. We start from the Hilbert space
$
{\cal H}^{(R)}_{\rm cov}
$
generated by the operators 
$
\alpha^{\mu}_{m}, m \in \Z, \mu = 0,\dots,D
$
and
$
d^{\mu}_{m}, m \in \Z, \mu = 0,\dots,D
$
such that:
\bea
[ \alpha^{\mu}_{m}, \alpha^{\nu}_{n} ] = - \eta_{\mu\nu}~m~\delta_{m+n} \cdot I, 
\quad \forall m, n
\nonumber \\
\alpha^{\mu}_{m} \Omega = 0, \quad m > 0
\nonumber \\
(\alpha^{\mu}_{m})^{+} = \alpha^{\mu}_{-m} \quad \forall m;
\nonumber \\
\{ d^{\mu}_{m}, d^{\nu}_{n} \} = - \eta_{\mu\nu}~m~\delta_{m+n} \cdot I, 
\quad \forall m, n
\nonumber \\
d^{\mu}_{m} \Omega = 0, \quad m > 0
\nonumber \\
(d^{\mu}_{m})^{+} = d^{\mu}_{-m} \quad \forall m
\eea
and define the covariant Virasoro operators
\be
L^{(R)}_{m} \equiv - {1\over 2}~\eta_{\mu\nu}~\sum_{n \in \Z}~
:\alpha^{\mu}_{m-n} \alpha^{\nu}_{n}:
- {1\over 2}~\eta_{\mu\nu}~\sum_{n \in \Z}~n~:d^{\mu}_{-n} d^{\nu}_{m+n}: 
- a~\delta_{m} \cdot I
\label{vir-R}
\ee
and the supersymmetric partners
\be
F_{m} \equiv - \eta_{\mu\nu}~\sum_{n \in \Z} \alpha^{\mu}_{-n} d^{\nu}_{n+m}
\ee
such that we have the following relations:
\bea
[ L^{(R)}_{m}, L^{(R)}_{n} ] = (m - n) L^{(R)}_{m+n} 
+ m \left( { D \over 8} m^{2} + 2 m a \right)~\delta_{m+n}~\cdot I. 
\nonumber \\
~[ L^{(R)}_{m}, F_{n} ] = \left({m\over 2} - n\right) F_{m+n}
\nonumber \\
\{ F_{m}, F_{n} \} = 2 L^{(R)}_{m+n} 
+ \left( {D\over 2}~m^{2} + 2 a \right)~\delta_{m+n}~\cdot I
\nonumber \\
(L^{(R)}_{m})^{\dagger} = L^{(R)}_{-m} \qquad F_{m}^{\dagger} = F_{-m}.
\eea
In this Hilbert space we have an action of the supersymmetric Poincar\'e algebra. Then we can obtain the Ramond case if we take 
$D = 10, a = 0$
and restrict the states by the conditions:
\be
L^{(R)}_{m}\Psi = 0 \quad \forall m \geq 0 \qquad
F_{m} \Psi = 0 \quad \forall m \geq 0.
\ee

The DDF states can be constructed as before. First we have to construct operators
$
A^{j}_{m},~D^{j}_{m}, j=1,\dots,D-1
$
such that they verify the same algebra as the operators
$
\alpha^{j}_{m}, d^{j}_{m}
$
and commute with
$
F_{m}, \forall m
$
(this implies that they commute with
$
L_{m}, \forall m.)
$
The DDF states are generated by these operators from the vacuum.

For the BRST description we need to enlarge the ghost space: we consider the Fock space
$
{\cal F}^{gh}_{3}
$
generated by the operators
$
\beta_{m}, \gamma_{m} \quad m \in \Z
$ 
from the vacuum
$\Omega_{gh} \in {\cal F}^{gh}_{3}$;
we assume that
\be
\beta_{m}\Omega_{gh} = 0 \quad \gamma_{m}\Omega_{gh} = 0 \quad \forall m > 0.
\ee

These operators are subject to the following anticommutation relations:
\be
[\beta_{m}, \beta_{n} ] = 0 \quad [ \gamma_{m}, \gamma_{n} ] = 0 \quad 
[ \gamma_{m}, b_{n} ] = \delta_{m+n} \cdot I;
\ee 
we also suppose that there is a conjugation operation in
$
{\cal F}^{gh}_{3}
$
such that
\be
\beta_{m}^{\dagger} = \beta_{-m} \quad \gamma_{m}^{\dagger} = \gamma_{-m}.
\ee

We can define as usual the algebraic Hilbert space (the subspace of vectors generated by a finite number of operators
$
\beta_{m}, \gamma_{m} \quad m \leq 0)
$ 
and normal ordering in
$
{\cal F}^{gh}_{m}.
$
\begin{prop}
The following operators 
\be
l^{(3)}_{m} = \sum_{n \in \Z} \left({m\over 2}+n\right) :\beta_{m-n}\gamma_{n}:
\ee
are well defined on the algebraic Hilbert space and are verifying:
\bea
[ l^{(3)}_{m}, \beta_{n}] = \left({m\over 2}-n\right) \beta_{m+n} \qquad
[ l^{(3)}_{m}, \gamma_{n}] = - \left({3m\over 2}+n\right) \gamma_{m+n}
\nonumber \\
~[ l^{(3)}_{m}, l^{(3)}_{n}] = (m-n) l^{(3)}_{m+n} 
+ {1\over 12} m(11m^{2} - 2) \delta_{m+n} \cdot I
\nonumber \\
(l^{(3)}_{m})^{\dagger} = l^{(3)}_{-m}.
\eea
\end{prop}
{\bf Proof:} We proceed as in Section \ref{brst}: first we split
\be
l^{(3)}_{m} = \tilde{l}^{\prime}_{m} + {m\over 2} \beta_{m}\gamma_{0} 
+ {3m\over 2} \beta_{0}\gamma_{m} 
\ee 
where the first term includes only the non-zero modes. For these modes the $2$-point functions are
\bea
<\Omega_{gh},\beta_{m}\gamma_{n}\Omega_{gh}> = -\theta(m) \delta_{m+n} \quad
<\Omega_{gh},\gamma_{m}\beta_{n}\Omega_{gh}> = \theta(m) \delta_{m+n} \quad
\nonumber \\
<\Omega_{gh},\beta_{m}\beta_{n}\Omega_{gh}> = 0 \quad
<\Omega_{gh},\gamma_{m}\gamma_{n}\Omega_{gh}> = 0
\eea
and we can compute the commutators from the statement using Wick theorem.
$\qed$

Next we have
\begin{cor}
Let us consider in the Hilbert spaces
$
{\cal F}^{gh}_{R} \equiv {\cal F}^{gh}_{1} \otimes {\cal F}^{gh}_{3}
$
the following operators 
\be
l^{(R)}_{m} = l^{(1)}_{m} \otimes I_{2} + I_{1} \otimes l^{(3)}_{m};
\ee
then we have
\bea
[ l_{m}^{(R)}, l_{n}^{(R)}] = (m-n) l_{m+n}^{(NS)}
- {5\over 4}~m^{3}~\delta_{m+n} \cdot I
\nonumber \\
(l_{m}^{(R)})^{\dagger} = l_{-m}^{(R)}.
\eea 
\end{cor}

Next, we define in
$
{\cal F}^{gh}_{R}
$
the following operators
\be
f_{m} \equiv - 2 \sum_{n \in \Z} b_{-n} \gamma_{n+m}
+ \sum_{n \in \Z} \left( {n\over 2} - m\right) c_{-n} \beta_{n+m}
\ee
which are well-defined on the algebraic Fock space. We have
\begin{prop}
The following relations are verified 
\bea
f_{m}^{\dagger} = f_{-m}
\nonumber \\
~[l_{m}^{(R)}, f_{n} ] = \left( {m\over 2} - n\right) f_{m+n}
\nonumber \\
\{ f_{m},f_{n}\} = 2 l_{m+n}^{(R)} - 5m^{2}~\delta_{m+n} \cdot I.
\eea
\end{prop}

The proofs of the first two relations are elementary. For the last one we use Wick theorem.
Next we have
\begin{cor}
Let us consider in the Hilbert spaces
$
{\cal H} \equiv {\cal H}^{(b)}_{cov} \otimes^{s} {\cal F}^{gh}_{R}
$
where the skew tensor product
$
\otimes^{s}
$
is such that we have normal (anti)commutation relations i.e. the Fermionic operators
$
d^{\mu}_{m}
$
are anticommuting with
$
b_{m}, c_{m}.
$
We then define the following operators 
\bea
{\cal L}_{m}^{(R)} = L^{(R)}_{m} \otimes I_{2} + I_{1} \otimes l^{(R)}_{m}
\nonumber \\
{\cal F}_{m} = F_{m} \otimes I_{2} + I_{1} \otimes f_{m}
\eea
then we have
\bea
[ {\cal L}_{m}^{(R)}, {\cal L}_{n}^{(R)}] = (m-n) {\cal L}_{m+n}^{(R)}
+ m \left( {D-10\over 8} m^{2} + 2a \right) \delta_{m+n} \cdot I
\nonumber \\
~[ {\cal L}_{m}^{(R)}, {\cal F}_{n}] = \left({m\over 2} - n\right) {\cal F}_{m+n}
\nonumber \\
\{ {\cal F}_{m}, {\cal F}_{n} \} = 2 {\cal L}^{(R)}_{m+n} 
+ \left( {D-10\over 8}~m^{2} + 2 a\right)~\delta_{r+s}~\cdot I
\nonumber \\
({\cal L}^{(R)}_{m})^{\dagger} = {\cal L}^{(R)}_{-m} \qquad 
{\cal F}_{m}^{\dagger} = {\cal F}_{-m}.
\eea 
The anomalies cancel iff
$D = 10, a = 0$.
\end{cor}
In this enlarged Hilbert space we have:
\begin{prop}
The following operator
\bea
Q = Q^{(R)} \equiv \sum L^{(R)}_{-m} c_{m} 
- {1\over 2} \sum (m-n) :c_{-m} c_{-n} b_{m+n}:
\nonumber \\
+ \sum F_{-m} \gamma_{m} + \sum c_{-m} l^{(3)}_{m}  + \sum b_{-m} C^{(4)}_{m}  
\eea
where
\be
C^{(4)}_{m} \equiv - \sum_{p+q=m} :\gamma_{p}\gamma_{q}: = 
- \sum_{p+q=m} \gamma_{p}\gamma_{q}
\ee
is well defined on the algebraic Hilbert space and is formally self-adjoint; it verifies
\be
Q^{2} = 0
\ee
{\it iff}
$
D = 10
$
and
$a = 0$.
\end{prop}
{\bf Proof:} It is convenient to denote by
$
Q_{j}, j = 1,\dots,5
$
the five terms in the expression of the BRST charge and write
\be
Q = Q^{\prime} + \sum_{j=3}^{5} Q_{j}
\ee
where the sum of the first two terms
$ 
Q^{\prime}
$
can be obtained from $Q$ of the preceding Section with the substitution
$
L^{(\alpha)}_{m} \longrightarrow L^{(R)}_{m}
$
so we can use some of the computations performed there. We introduce the notation
\be
H_{m} \equiv k_{\mu}~d^{\mu}_{m}. 
\ee
and we have as before:
\bea
\{Q, b_{m} \} = {\cal L}_{m}^{(R)} \qquad
\{ Q, c_{m} \} = C^{(R)}_{m} \equiv C^{(1)}_{m} + C^{(4)}_{m}
\nonumber \\
~[ Q, \beta_{m} ] = {\cal F}_{m} \qquad
[ Q, \gamma_{m} \} = - \sum \left( {3n\over 2} + m\right) c_{-n} \gamma_{m+n}
\nonumber \\
~[ Q, {\cal L}_{m}^{(R)} ] = \rho_{m} c_{m} \qquad
\{ Q^{(R)}, {\cal F}_{m} \} = \lambda_{m} \gamma_{m}
\nonumber \\
~[ Q, K_{m} ] = - m \sum K_{m-n} c_{n} + \sum H_{m-n} \gamma_{n}
\nonumber \\
\{ Q, H_{m} \} = - \sum \left( {n\over 2} - m\right) H_{m-n} c_{m} + \sum K_{m-n} \gamma_{n}.
\label{Q-r1}
\eea
where
\be
\rho_{m} \equiv - m \left( {D-10\over 8} m^{2} + 2a \right)
\qquad 
\lambda_{m} \equiv {D-10\over 2} m^{2} + 2a.
\ee

One can prove that we can take in
${\cal H}$
the following basis:
\be
\Psi = \prod b_{-i} \prod c_{-j} \prod \beta_{-p} \prod \gamma_{-q}
\prod {\cal L}_{-m}^{(NS)} \prod {\cal F}_{-l} \prod K_{-n} \prod H_{-k}~f
\label{basis-r}
\ee
where $f$ are DDF states, the indices of the type $m,n \in \Z$
are taking positive values and the indices of the type $i,j, p, q, k, l \in \Z$ are $\geq 0$.

Because
$
L_{m}^{(R)} f = 0 \quad {\cal F}_{m} f = 0 \quad \forall m \geq 0
$
we easily find out that
\be
Qf = 0
\label{Q-r2}
\ee
for any DDF state $f$. We argue as before that the operator 
$Q$ 
is well defined by (\ref{Q-r1}) and (\ref{Q-r2}). Now it is easy to obtain from (\ref{Q-r1}) that
$
Q^{2}
$
commutes with all the operators from (\ref{basis-r}). Because we have
$
Q^{2}f = 0
$
it immediately follows that
\be
Q^{2} = 0 \Longleftrightarrow \rho^{(R)}_{m} = 0 \quad \lambda^{(R)}_{m} = 0 \Longleftrightarrow D = 10, a = 0 
\label{Q-r5}
\ee 
i.e. the statement of the theorem.
$\qed$

To analyze the cohomology of the BRST operator 
$Q$ 
we construct its homotopy:
\begin{prop}
The operator 
$\tilde{Q}$
is well defined on the algebraic Hilbert space through the following formulas:
\bea
\{\tilde{Q}, b_{m} \} = 0 \qquad
\{ \tilde{Q}, c_{m} \} = \delta_{m} \cdot I \qquad
\{\tilde{Q}, \beta_{m} \} = 0 \qquad
\nonumber \\
\{ \tilde{Q}, \gamma_{m} \} = 0 \qquad
~[ \tilde{Q}, {\cal L}_{m}^{(R)} ] = - m b _{m} \qquad
\{ \tilde{Q}, {\cal F}_{m} \} = - m \beta_{m}
\nonumber \\
~[ \tilde{Q}, K_{m} ] = 0 \qquad
[ \tilde{Q}, H_{m} ] = 0
\label{tQ-r1}
\eea
and
\be
\tilde{Q}f = 0
\ee
for any DDF state $f$. We also have
\be
\tilde{Q}^{\dagger} = \tilde{Q} \qquad \tilde{Q}^{2} = 0.
\ee
\end{prop}

{\bf Proof:} We have to verify the Jacobi identities of the type:
\be
[[X,Y], \tilde{Q} ]_{\rm graded} + {\rm cyclic~permutations} = 0
\ee
where
$X, Y$
are operators from the set
$
b_{m}, c_{m}, \beta_{m}, \gamma_{m}, {\cal L}_{m}^{(R)}, {\cal F}_{m}, K_{m}, H_{m}.
$ 
We have some non-trivial ones corresponding to pairs
$
({\cal L}_{m}^{(R)}, {\cal L}_{n}^{(R)}),~
({\cal L}_{m}^{(R)}, c_{n}),~
({\cal L}_{m}^{(R)}, {\cal F}_{r}),
({\cal F}_{m}, {\cal F}_{n})
$
and
$
({\cal F}_{m}, \gamma_{m}).
$
$\qed$

The main result is similar to the one in the previous Section. However, the degeneracy is infinite
in this case, so to avoid this problem we work directly in a smaller Hibert space: we construct the Fock space 
${\cal F}^{gh}_{R}$
such that it also verifies 
\be
b_{0} \Omega_{gh} = 0 \qquad \beta_{0} \Omega_{gh} = 0.
\ee
Then we construct ${\cal H}$ as above and consider the subspace
\be
{\cal H}_{0} \equiv \{ \Psi \in {\cal H} | \quad b_{0}\Psi = 0 \quad \beta_{0} \Psi = 0 \quad {\cal L}^{(R)}_{0} \Psi = 0 \quad {\cal F}_{0} \Psi = 0\}.
\ee

This subspace is left invariant by the operator 
$Q^{(R)}$
and we have the following result:
\begin{thm}
If 
$
\Psi \in {\cal H}_{0}
$
verifies
$
Q\Psi = 0
$
then it is of the form 
\be
\Psi = Q\Phi + f
\ee
where 
$
f
$
is a DDF state and
$
\Phi \in {\cal H}_{0}.
$
\end{thm}
{\bf Proof:} 
First we have to prove that the states from
$
{\cal H}_{0}
$
are obtained applying the operators
$
b_{-m}, c_{-m}, \beta_{-m}, \gamma_{-m}, {\cal L}^{(R)}_{-m}, {\cal F}_{-m}, K_{-m}, H_{-m} 
$
with
$m > 0$.

The ``Laplacian" is as before
$
\Delta \equiv Q\tilde{Q} + \tilde{Q}Q
$
It is now elementary to determine that the Laplace operator is alternatively given by:
\bea
[ \Delta, b_{m} ] = - m b_{m} \quad [ \Delta, c_{m} ] = - m c_{m} \qquad
[ \Delta, \beta_{m} ] = - m \beta_{m} \quad 
[ \Delta, \gamma_{m} ] = - m \gamma_{m} 
\nonumber \\
~[ \Delta, {\cal L}^{(R)}_{m} ] = - m {\cal L}^{(R)}_{m} \quad 
[ \Delta, {\cal F}_{m} ] = - m {\cal F}_{m} \qquad
[ \Delta, K_{m} ] = - m K_{m} \quad
[ \Delta, H_{m} ] = - m H_{m}
\nonumber \\
\Delta f = 0.
\eea

It follows that we have
\be
\Delta \Psi = (\sum m + \sum n + \sum i + \sum j + \sum p + \sum q + \sum k + \sum l ) \Psi;
\ee 
we observe that the eigenvalues from (\ref{delta}) are 
$\geq 0$
the equality sign being true only for DDF states. From now on the argument is the same as before.
$\qed$

\section{Conclusions}
The main results of this paper are: a) An elementary treatment of the quantum string models relying only on Wick theorem and paying attention to the domain problems. b) A derivation of the DDF operators without using vertex algebras. c) The clarification of the equivalence between the light-cone and covariant formalism using standard results in induced representation theory; this point seems to be missing from the literature. d) An elementary derivation of the BRST cohomology. A comparison with the standard literature is useful on this point: In \cite{T} one uses a basis of the type (\ref{basis-gh}):
\be
\Psi_{I,J,M,N} = b_{-i_{1}} \dots b_{-i_{\beta}} c_{-j_{1}} \dots c_{-j_{\gamma}}
{\cal L}_{-m_{1}} \cdots {\cal L}_{-m_{\lambda}} 
K_{-n_{1}} \cdots K_{-n_{\kappa}}~f_{I,J,M,N}
\label{basis-ghT}
\ee
where it can be arranged such that the DDF states 
$
f_{I,J,M,N}
$ 
are completely symmetric in the indices
$
M = \{ m_{1},\dots, m_{\lambda}\}
$
and in the indices
$
N = \{ n_{1},\dots, n_{\kappa}\};
$
of course we have complete antisymmetry in the indices
$
I = \{ i_{1},\dots, i_{\beta}\}
$
and in the indices
$
J = \{ j_{1},\dots, j_{\gamma}\}.
$
Then one decomposes the f's according to Young diagrams (separately for 
$
I \cup M
$
and
$
J \cup N).
$
We have in both cases only two projectors: one piece is eliminated by the condition
$Q\Psi = 0$
and the other one can be proved to by a coboundary up to states of the form
$
f_{1} + b_{0} f_{2}.
$
In \cite{FGZ} there are two proofs, one based on a similar idea of Hodge theory (however the expression of the Laplacian seems to be different and the spectral analysis is not provided) and the other proof relies on the use of spectral sequences.  

The proof from \cite{KO}, \cite{Oh} makes a convenient rescaling by a parameter
$\beta$ 
and assumes that the states  
$\Psi(\beta)$ 
are polynomials in this parameter which is an unjustified restriction. The proof from
\cite{P} is closely related and assumes that a certain infinite series
is convergent. The proof from \cite{FB} relies on the existence of the
operators
$
D_{n}
$
formally given by:
$
:\left(1 - \sum K_{m} z^{m}\right)^{-1}: = \sum z^{n} D_{n};
$
such operators are also used in the construction of the DDF states for the superstring models.


\end{document}